# On the Modeling and Performance Assessment of Random Access with SIC

Alberto Mengali, Riccardo De Gaudenzi, Senior Member, IEEE, Čedomir Stefanović, Senior Member, IEEE

***Abstract*—In this paper, we review the key figures of merit to assess the performance of advanced random access (RA) schemes exploiting physical layer coding, repetitions and collision resolution techniques. We then investigate RA modeling aspects and their impact on the figures of merit for the exemplary advanced RA schemes: Contention Resolution Diversity Slotted ALOHA (CRDSA), Irregular Repetition Slotted ALOHA (IRSA), Coded Slotted ALOHA (CSA) and Enhanced Spread-Spectrum ALOHA (E-SSA). We show that typical simplifications of the reception model when used to optimize RA schemes lead to inaccurate findings, both in terms of parameter optimization and figures of merit, such as the packet loss ratio (PLR) and throughput. We also derive a generic RA energy efficiency model able to compare the schemes in terms of the energy required to transmit a packet. The combination of achievable RA throughput at the target PLR and energy efficiency, for the same average user power investment per frame and occupied bandwidth, shows that E-SSA, which is an unslotted scheme, provides the best overall performance, while, in terms of the slotted schemes, CRDSA outperforms the more elaborated IRSA and CSA. This surprising results is due to the fact that the IRSA and CSA optimization has so far been performed using RA channel models that are not accurately reflecting the physical layer receiver behavior. We conclude by providing insights on how to include more accurate reception models in the IRSA and CSA design and optimization.**

## I. Introduction and Motivation

Recent years have witnessed a strong interest in Machine-to Machine (M2M) communications, i.e., communications among devices without or with minimal human intervention. Examples of M2M services include security, tracking, payment, smart grid and remote devices' maintenance and monitoring. M2M communication is a key enabler of the Internet of Things (IoT), where the central challenges are related to device energy efficiency and network scalability [1], [2]. Specifically, in relation to the latter, M2M scenarios often involve a large number of devices, where random subsets of devices become activated at a time and access the network in an uncoordinated manner. This has revamped the interest in research on Random Access (RA) schemes applicable to ground and space wireless systems.

In the scenarios involving M2M communications with satellites, the set of M2M devices within a satellite coverage can be vast, which calls for new designs of RA schemes. It has been shown that the adoption of relatively simple Successive Interference Cancellation (SIC) techniques makes it possible to achieve huge RA throughput performance improvements. An extensive survey of this kind of RA schemes in the context of satellite communications can be found in [3]. In general, RA schemes with SIC application is not confined to satellite systems. We refer the interested reader for an overview of slotted ALOHA with SIC [4], which establishes analogies with codes-on-graphs, as well as to some recent results presented in [5], [6].

The performance assessment of RA schemes is usually performed using simplified models of the physical layer (PHY) reception, where de facto standards are the collision-channel model and the threshold-based model. Although these models enable elegant analytical treatment and optimization of RA schemes with SIC, as shown in [7], they often lead to unreliable performance assessment. This insight calls for the use of an accurate methodology for performance assessment, such that emerging RA schemes with SIC can be compared in a fair and realistic way, which is the main motivation of this paper.

Specifically, in this paper, we assess the performance of the exemplary high-performing RA schemes with SIC, namely Contention Resolution Diversity Slotted ALOHA (CRDSA) [8], Irregular Repetition Slotted ALOHA (IRSA) [9], Coded Slotted ALOHA (CSA) [10] and Enhanced Spread-Spectrum ALOHA (E-SSA) [11], using the collision-channel model, the threshold-based model, and an accurate, semi-analytical (SEA) model of the PHY-layer receiver processing. We show that the simplified models are indeed inadequate to capture the particularities of PHY-layer processing. Instead, the use of the collision channel model provides rather inaccurate results, while the threshold-based model, even with a careful identification of the threshold value (that depends on the parameters of PHY layer), has only a limited applicability.

Another distinguishing feature of the paper is that the presented performance assessment deals with both the arrivals with constant number of users and with Poisson arrivals. While the former model is standardly used in literature on RA schemes with SIC, the latter is more appropriate to model periodic reporting in M2M services. In this regard, we

A. Mengali is with University of Luxembourg, Interdisciplinary Centre for Security, Reliability and Trust, 4, rue Alphonse Weicker L-2721 Luxembourg; R. De Gaudenzi is with European Space Agency, Keplerlaan 1, 2200 AG Noordwijk, The Netherlands; Č. Stefanović is with the Department of Electronic Systems, Aalborg University (Copenhagen campus) Fredrikskaj 12, 2450 Copenhagen, Denmark. e-mail: {rdegaude@gmail.com, a.mengali@gmail.com,cs@cmi.aau.dk}.

The work of Č. Stefanović was supported in part by the European Research Council (ERC Consolidator Grant Nr. 648382 WILLOW) in the Horizon 2020 Program. The work from A. Mengali was in part supported by the National Research Fund, Luxembourg under AFR grant for Ph.D. project (Reference 10064089) on Reliable Communication Techniques for Future Generation Satellite Systems.

show that the arrival model leaves its mark on the scheme performance.

The rest of the paper is organized as follows. Sec. II provides a brief introduction to the RA schemes with SIC investigated in the paper. Sec. III describes the key RA parameters and figures of merit that will be used, Sec. IV describes specific RA modeling aspects that are relevant for the performance assessment. In Sec. V, RA performance results based on simplified reception models are compared to the ones obtained with a substantially more accurate physical layer modeling, including an overall comparison in terms of normalized throughput and energy efficiency. Finally, Sec. VI provides recommendations based on lessons learned to guide future research work.

## II. ALOHA protocols with Successive Interference Cancellation

CRDSA, IRSA and CSA belong to the class of framed slotted ALOHA schemes [12], where the MAC-layer frame[1] consists of a fixed number of slots. In CRDSA [8], each user transmits *a fixed* number of physical-layer replicas of its data packet in independently and randomly selected slots of the frame. Each replica contains in the header information about the location of all other related replicas[2]. Hence, a successful decoding of a replica enables removal (i.e., cancellation) of all the other replicas of the packet from the slots in which they occur, which helps the demodulator to decode new packets (i.e., their replicas) from the affected slots.

IRSA represents a generalization of CRDSA, characterized by a random number of transmitted replicas per user. The number of replicas is selected according to a probability mass function (pmf) that is optimized off line and common to all users.

The basic idea of CSA is to transmit encoded segments of data packets, instead of transmitting their replicas. Specifically, a data packet that is about to be transmitted in a frame is divided into $k$ packet segments. The $k$ packet segments are then encoded via a segment-oriented binary linear block code that generates $n_h$ encoded segments, where the employed code is drawn randomly by the user from a finite set of codes. At the receiver side, successful decoding of any subset of $k$ independent segments enables the recovery of the whole data packet, as well as removal of the rest of the segments via interference cancellation. IRSA may be seen as a special case of CSA where $k = 1$ and each block code is a repetition code.

E-SSA is an unslotted (i.e., pure) ALOHA-based RA scheme in which packets are coded, modulated, spread and transmitted asynchronously at the same carrier frequency without any repetition and coordination. Collisions are resolved by means of SIC that exploits Direct-Sequence Spread-Spectrum (DS-SS) techniques to decorrelate the colliding packets.

## III. RA Figures of Merit

In order to provide a fair comparison, we assume that the occupied bandwidth per frame is the same for all RA schemes under consideration. We also assume that the average transmission power per packet replica over the frame $\overline{P}_f$ is the same for all considered RA schemes. All the schemes under consideration use a common fixed-length frame duration $\tau_f$.

In CRDSA and IRSA, a frame is composed of $N_{\text{slots}}$ slots, where each slot fits a packet transmission, and all users are slot-synchronized. A user sends $N_{\text{rep}}$ packet replicas, $N_{\text{rep}} \leq N_{\text{slots}}$, in randomly selected slots. For CRDSA, $N_{\text{rep}}$ is fixed, while for IRSA $N_{\text{rep}}$ can vary on the user basis and we denote by $\overline{N}_{\text{rep}}$ the average number of replicas sent per user.

In CSA, a user sends encoded segments in randomly chosen *slices* of the frame. The relation between slices in CSA and slots in CRDSA/IRSA can be easily established: as the size of the segment is $k$ times shorter than the size of the packet (see Section II), this implies that there are $kN_{\text{slots}}$ slices in CSA. The number of transmitted encoded segments in CSA is $n_h$, where $n_h$ can vary on user basis. Thus, in CSA, a user effectively transmits $N_{\text{rep}} = \frac{n_h}{k}$ packets in the frame, and the average number of transmitted replicas is $\overline{N}_{\text{rep}} = \frac{\overline{n}_h}{k}$.

In E-SSA there are no slots, as the packet is occupying the whole frame duration, thus, $N_{\text{slots}} = 1$, and there is no frame synchronization, i.e., frame references are user specific. Further, there are no distinct repetitions of the physical layer packet as in CRDSA and IRSA, but the packet symbols are repeated through spreading, which is accounted for through a spreading factor $\text{SF} > 1$. Thus, $N_{\text{rep}} = 1$ and a user packet is sent during the whole duration of the user frame.

We continue by the definition and justification of the key RA figures of merit adopted in the paper. In Appendix D of [13], it was shown that, assuming the packets are generated according to a Poisson distribution, the aggregate traffic (over all users and slots) mean data packet arrival rate per frame $\lambda^f_{\text{tot}}$ can be computed as

$$\lambda^f_{\text{tot}} = N_{\text{slot}}\, G\, G_p, \tag{1}$$

where $G$ is the MAC load expressed in data bits/symbol in non-SS systems and bits/chip in SS systems, and where $G_p$ is the processing gain, defined as the ratio between the chip rate $R_c$ expressed in chips/s[3] and the effective data bit rate $R_b$ during transmission, given by

$$G_p = \frac{R_c}{R_b} = \frac{1}{r^* \log_2 M}, \tag{2}$$

where $r^* = r/\text{SF}$, $r$ being the Forward Error Correction (FEC) code rate, SF is the spreading factor (if applicable) and $M$ is the modulation cardinality. In (2), we assume that the spreading corresponds to repetition coding with rate $1/\text{SF}$ concatenated to the FEC.

Recall that, in time slotted RA schemes, the frame duration is divided in $N_{\text{slots}}$ over which packets are transmitted. Thus, the data packet bit rate averaged over the frame duration is simply given by

$$[\overline{R}_b]_f = \frac{R_b}{N_{\text{slots}}}. \tag{3}$$

[1] Later on referred for simplicity as frame.

[2] In principle, this information can be agreed in advance through, e.g., a shared seed for a random number generator.

[3] In the following, we use the general term chip which is typical of spread-spectrum (SS) systems. Clearly, for non SS RA schemes, i.e., when $\text{SF} = 1$, the chip correspond to the symbol and the chip rate to the baud rate.

*1) Occupied Bandwidth:* Following (2) and (3), the RA scheme occupied bandwidth, i.e., the signaling rate is given by

$$R_c = R_b G_p = [\overline{R}_b]_f G_p N_{\text{slots}}. \tag{4}$$

Specializing (4) to the case of slotted non SS and unslotted SS RA schemes we get

$$R_c = \frac{[\overline{R}_b]_f N_{\text{slots}}}{r^* \log_2 M}. \tag{5}$$

From (5), it is evident that, if we want to transmit the same average bit rate over the RA frame having the *same time duration* and occupying the *same bandwidth*, we should ensure that the quantity $\frac{N_{\text{slots}}}{r \log_2 M}$ for slotted non SS RA has the same value as $\frac{\text{SF}}{r \log_2 M}$ for an unslotted SS scheme. For the case of CRDSA, the preferred values are $M = 4$ and $r = 1/3$ [13], while for E-SSA $M = 2$ and $r = 1/3$ [11]. Thus, for a fair comparison between E-SSA and CRDSA, we should assume $\text{SF} = N_{\text{slots}}/2$. This finding shows that the number of slots are expanding the RA scheme occupied bandwidth similarly to the spreading in a SS scheme.

### A. MAC Normalized Load

In general, the RA traffic is bursty and the instantaneous MAC load is thus a random variable (rv). Once the traffic model has been defined, the MAC load can be characterized by its average value. Following (1), the average normalized MAC load expressed in bits/symbol (or bits/chip in case of SS RA) is computed as

$$G = \frac{\lambda_{\text{tot}}^{f}}{G_p N_{\text{slot}}}. \tag{6}$$

### B. Packet Loss

The Packet Loss Ratio (PLR) is defined as the ratio between the number of incorrectly decoded packets and the total number of data packets sent at media access control level. The PLR depends on the MAC average load $G$.

### C. MAC Normalized Throughput

The normalized throughput $T$ represents the spectral efficiency of the RA scheme at given average MAC load $G$ and is expressed in the same unit as $G$. The normalized throughput is related to the PLR through the following simple equation [13]

$$T(G) = G \cdot [1 - \text{PLR}(G)]. \tag{7}$$

### D. Energy Efficiency

The energy efficiency of a RA scheme is of paramount importance, since the devices in M2M applications are often operated with battery or solar based energy supply. We define the energy efficiency $\psi_e$ (bits/Joule) as the ratio between the average number of data bits $[N_b]_f$ successfully transfered by a user in a frame and the energy $E_f$ required to transmit them. Analytically, this can be expressed as

$$\psi_e = \frac{[N_b]_f}{E_f}. \tag{8}$$

The energy required to transmit the $[N_b]_f$ bits over the frame can be expressed as

$$E_f = \overline{P}_f \overline{N}_{\text{rep}} \tau_f. \tag{9}$$

The normalized throughput is then related to the average number of data bits successfully transmitted in a frame divided by the signaling rate and the frame duration. This can be expressed as

$$T(G) = \frac{\lambda_{\text{tot}}^{f}(G)\; [N_b]_f}{\tau_f R_c}. \tag{10}$$

Now, replacing(10) and (9) in (8), we obtain

$$\psi_e = \frac{T(G) R_c}{\overline{N}_{\text{rep}} \lambda_{\text{tot}}^{f} \overline{P}_f} = \psi_e^n \cdot \frac{R_c}{\overline{P}_f}, \tag{11}$$

where we exploit the definition of the normalized energy efficiency $\psi_e^n$, which is the energy efficiency divided by $R_c/\overline{P}_f$, previously assumed constant for comparison fairness. This assumption makes the comparison between different schemes dependent on the normalized efficiency only. In practice, the RA scheme is typically operated at $\text{PLR} < 10^{-2}$ to minimize retransmissions and to ensure stability. Thus, in deriving $\psi_e$ we will consider the average MAC load $G^*$ corresponding to a $\text{PLR}^* = 10^{-3}$. Following (7), it is clear that $T(G^*) \simeq G^*$ for the selected $\text{PLR}^*$ value. By taking all the previous approximations into account, exploiting (1) and (3), the normalized energy efficiency can be approximated as

$$\psi_e^n \approx \frac{1}{\overline{N}_{\text{rep}} N_{\text{slot}} G_p} = \frac{r \log_2 M}{\overline{N}_{\text{rep}} N_{\text{slot}}\ \text{SF}}, \tag{12}$$

where in the final formulation of $\psi_e^n$ we simply exploited (2).

## IV. Modeling Aspects

### A. Traffic Models

The simplest traffic model is the one in which the load on the frame basis is constant, allowing for straightforward optimization of the RA scheme. A more accurate model is the one that assumes Poisson distribution of the load on the frame basis, which also implies, for slotted schemes, a Poisson distribution of the load at slot level (the proof is presented in Appendix A). This is in line with the recommendations of the 3rd Generation Partnership Project (3GPP) for asynchronously reporting M2M traffic [14]. In scenarios in which devices access the network in a highly synchronized manner, 3GPP suggests using a traffic model based on beta distribution. In the following, we will focus on the case of asynchronous arrivals, typical for massive M2M services, and compare the impacts of Poisson distributed load and constant load on the performance.

An investigation about the impact of the RA scheme and physical layer parameters on the traffic probability distribution normalized to the Poisson distribution mean value has been reported in [15]. The conclusion is that the instantaneous number of interfering packets will fluctuate less, in proportion to its average, when $\lambda_{\text{tot}}^{f}$ increases. Recalling (1), it is evident that for a given average load $G$, $\lambda_{\text{tot}}^{f}$ is directly proportional to the number of slots in the frame (if applicable), the spreading factor (if applicable) and inversely proportional to the physical layer spectral efficiency ($r \log_2 M$). As an example, a RA scheme exploiting spread-spectrum with $\mathrm{SF} \gg 1$, experiences less interference fluctuation normalized to its average value than a non spread-spectrum RA scheme. Thus, as shown in Fig. 2 of [15], the demodulator should be designed to cope with a worst-case level of interference that is close to its average value and proportional to $\lambda_{\mathrm{tot}}^{f}$. Instead, in a non spread-spectrum RA, the demodulator should cope with worst-case interference levels which are proportional to about $3\lambda_{\mathrm{tot}}^{f}$.

### *B. Packets' Power Unbalance*

Practical systems often face a random distribution of the received packets' power. This is because path loss, antenna gain (for both transmitter and receiver) and channel propagation conditions are typically different for the different packets. Closed-loop power control should be avoided in satellite M2M systems to minimize the transmission time and the signaling overhead. Some systems, characterized by a two-way terminal communication capability, may implement a simpler open-loop power control to enhance the system throughput performance. For CRDSA and E-SSA this aspect has been investigated in [7], [13] and [11], [16], respectively. For both schemes, it has been found that packets power randomization following a uniform distribution in dB represents a close to optimum way to increase the RA throughput. The minimum incoming packet power level should be selected to make the packet decodable in the absence of collisions, while the maximum power level depends on the available transmitter power and system link budget. References [16], [17] show how this close-to-optimum received packets' power distribution can be obtained with a simple open-loop power control algorithm.

### *C. Modeling of Physical Layer Reception*

*1) Collision Model:* A typical model of PHY-layer reception assumed in the literature on ALOHA-based schemes is the collision channel model, cf. [12]. According to this model, all packets involved in a collision are lost, and packets that do not experience collisions are always successfully decoded. In reality, this approximate model does not hold when: (i) received powers of the packets are not the same, (ii) powerful FEC code is used, (iii) when DS-SS is adopted. In the first case, the colliding packet that has the highest power may still have a Signal-to-Noise plus Interference Ratio (SNIR) sufficient to be decoded. This is often referred to as *capture* effect. In the second case, if a Turbo or Low Density Parity Check (LDPC) FEC with a low code rate (e.g., 1/3) and Binary or Quaternary Phase Shift Keying (BPSK/QPSK) modulation are used, the demodulator might still be able to resolve collisions of cardinality 2 at physical layer level[4]. In the third case, all the packets involved in collisions whose cardinalities are up to the SF are inherently accepted by the system, i.e., the use of spreading enables multi-packet reception.

*2) Threshold-based Model:* Another frequently used model is the threshold-based reception model, which assumes that a received packet subject to collisions can be decoded with probability 1 if its SNIR is above a given threshold $\rho_{\mathrm{FEC}}$, and decoded with probability 0 if the SNIR is below, cf. [18], [19]. The threshold-based reception model attempts to incorporate the aspects outlined above through a single parameter - the value of the threshold, which is unsuitable for the scenarios pertaining to satellite communications. Specifically, as elaborated [7], [11], [13], the decoding probability depends on the physical layer characteristics and does not exhibit an abrupt transition from 0 to 1 as function of SNIR, as the threshold-based model suggests. Moreover, as shown in the following section, the results obtained with a simple threshold-based reception model heavily depend on the choice of the $\rho_{\mathrm{FEC}}$ value, and finding the right value of $\rho_{\mathrm{FEC}}$ that models appropriately the physical layer operation is not an easy task.

In summary, the preferred approach to analyze the performance of RA schemes in satellite communications seems to be resorting to the accurate simulation of the physical layer, thus including the actual modulation and coding scheme.

*3) Semi-Analytical Model:* To speed-up simulations, semi-analytical (SEA) models can be adopted [7], [11], [13]. The SEA approach is representing the iterative RA demodulator operation in the presence of a given packets traffic and power distribution. SEA models approximate the colliding packets as as an Additive White Gaussian (AWGN) process, which is on top of the thermal noise. Furthermore, the logarithmic packet-error performance of PHY layer (i.e., of the coding and modulation scheme) in presence of AWGN noise is fitted by a polynomial function.

In the following, for the accurate PHY model we refer to the one described in [7] for CRDSA, IRSA and CSA and the one reported in Sect. III.A of [11] for E-SSA. Their accuracy has been validated through extensive comparison with the results of Monte Carlo simulations of the full PHY-layer processing. For E-SSA, as the SEA results are not as accurate as for the other RA schemes, we also provide results with the Monte Carlo physical layer simulation approach described in Sect. III.D of [11].

## V. Performance Comparison

In this section, we assess the performance of CRDSA, IRSA, CSA and E-SSA schemes using the traffic and PHY-layer reception models described in the previous section.

### *A. CRDSA*

*1) Constant vs Poisson Traffic:* We first investigate the impact of Poisson distributed versus constant aggregate load for CRDSA with 2, 3, and 4 replicas, using the accurate semi-analytical PHY model and limiting the number of SIC iteration at the demodulator to $N_{\mathrm{iter}} = 20$. Results are shown in Fig. 1, where the difference between the two traffic models is clear from the PLR waterfall region. As expected, following Sec. IV-A, the throughput results for Poisson traffic model with the same average normalized load $G$ are worse than for the constant traffic, due to the presence of load peaks which negatively affect the demodulator performance. For 3 replicas at $\mathrm{PLR} = 10^{-3}$, the throughput loss corresponds to about 15 %. To provide realistic results, in the following we will use the model with Poisson traffic, which is also more conservative in terms of performance.

[4]By cardinality, we denote the number users/replicas involved in a collision.

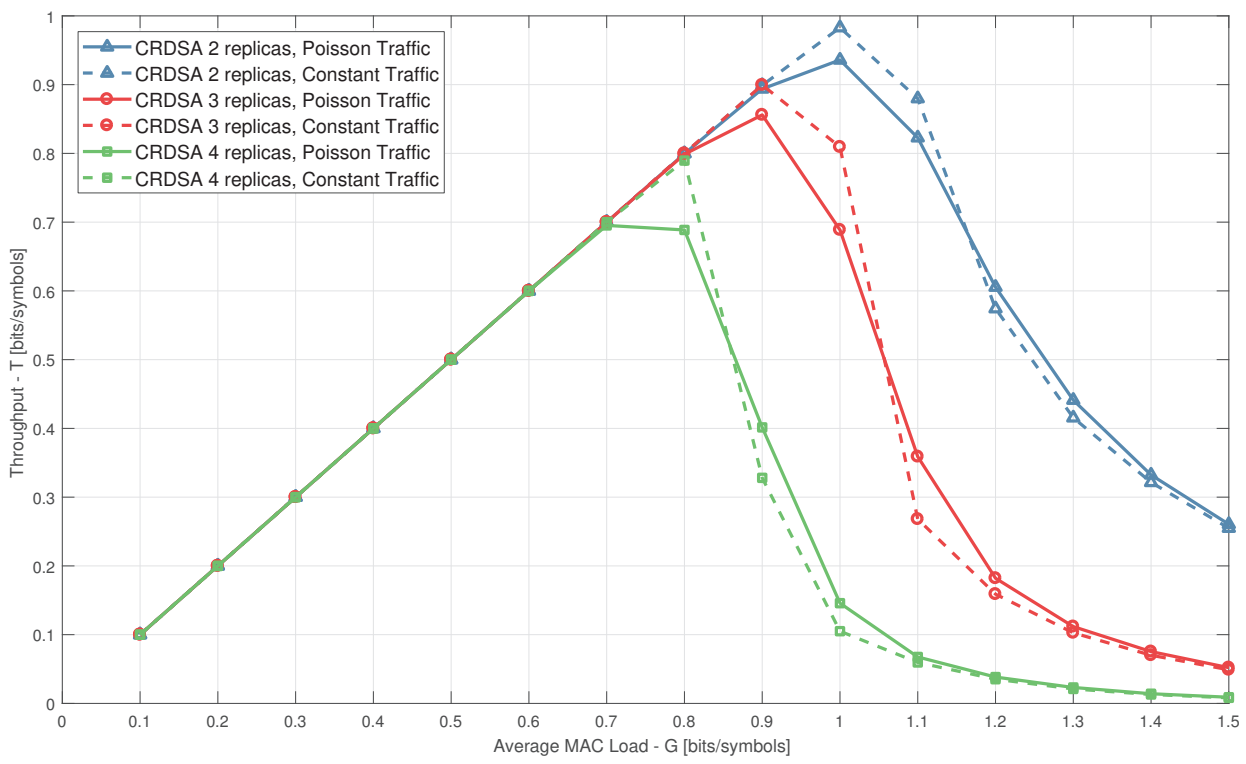

(a) Throughput

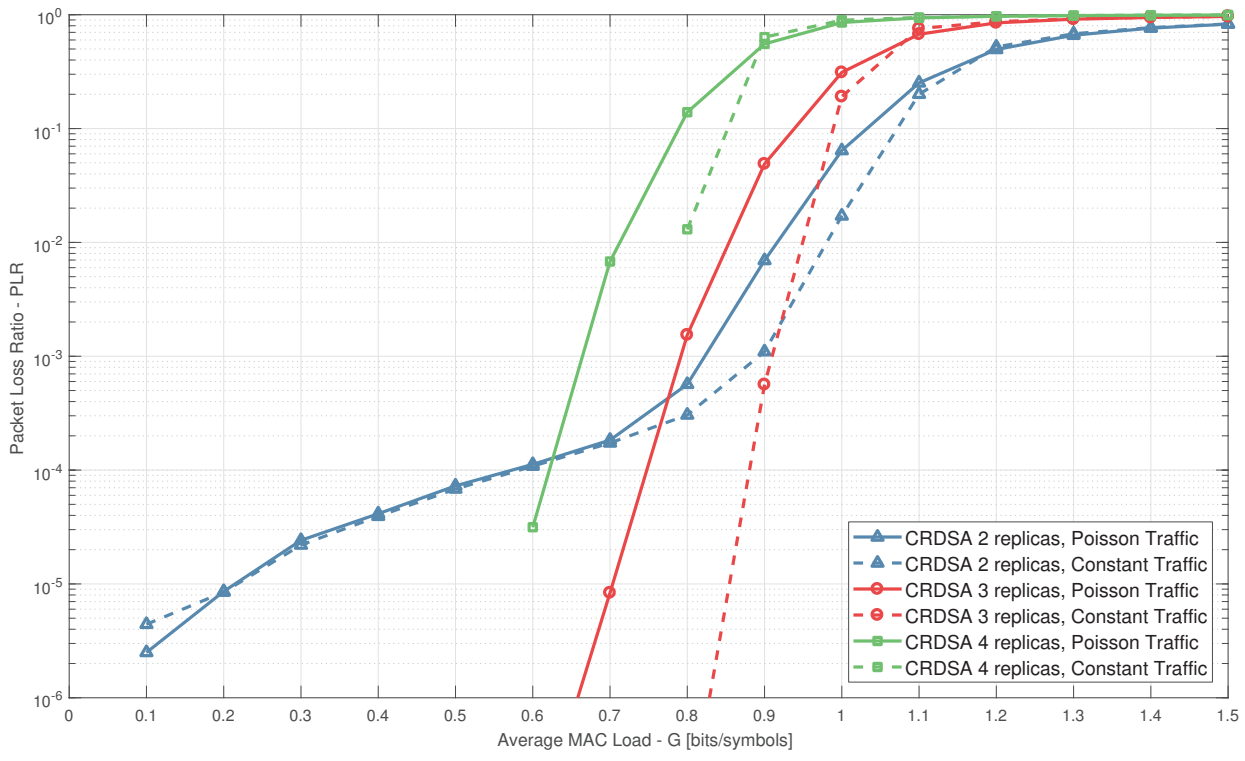

(b) PLR

Fig. 1. Comparison of simulated CRDSA throughput and PLR with Poisson and constant traffic as a function of the average load $G$ for 2, 3 and 4 replicas; accurate PHY model, $r = 1/3$ FEC, QPSK modulation, $N_{\text{slots}} = 128$, $N_{\text{iter}} = 20$, $E_s/N_0 = 10$ dB.

*2) Simple Reception Models vs Accurate PHY Model:* We now investigate the difference between the accurate PHY model and the collision channel model/threshold-based model described in Sec. IV-C for the 3GPP turbo FEC with rates 1/3 and 1/2 [20]. It is worth mentioning that in the case of equi-powered packets, there is a finite number of possible SNIR values. The SNIR that each packet is subject to is in fact uniquely identified by the system $E_s/N_0$ and by the cardinality of the collision experienced by the packet. For example, when $E_s/N_0 = 10$ dB, the SNIR of a packet involved in a collision of cardinality 2 is $\text{SNIR}_2 = \frac{E_s}{E_s+N_0} = -0.41$ dB, while the equivalent SNIR of a packet in a collision of cardinality 3 is $\text{SNIR}_3 = \frac{E_s}{2E_s+N_0} = -3.22$ dB. A simple reasoning reveals that, when $E_s/N_0 = 10$ dB, if $\rho_{\text{FEC}} > -0.41$ dB, the threshold-based model becomes equivalent to the collision channel model. Similarly, when $-3.22$ dB $< \rho_{\text{FEC}} < -0.41$ dB, the threshold-based reception model always resolves collisions of up to cardinality 2.

The results comparing these simplified models with the accurate one for CRDSA with 2, 3 replicas and FEC rates 1/2 and 1/3 are presented in Fig. 2. To avoid excessive clutter in the legends, in this figure and in the following ones, we will use 'P.M.', 'C.M.' and 'T.M.' to represent the accurate PHY model, collision channel model and threshold-based reception model, respectively. While for $r = 1/3$, the threshold-based model seems to give a fairly accurate match with the accurate PHY model results, the results for $r = 1/2$ show that a match is not achievable. This result can be explained by the fact that, for $E_s/N_0 = 10$ dB, a packet employing QPSK modulation and the 3GPP FEC code with $r = 1/2$ has a low but non-negligible probability of being correctly decoded when experiencing a collision of cardinality 2. This makes the collision channel slightly pessimistic and the threshold model with $\rho_{\text{FEC}} = -2$ dB too optimistic

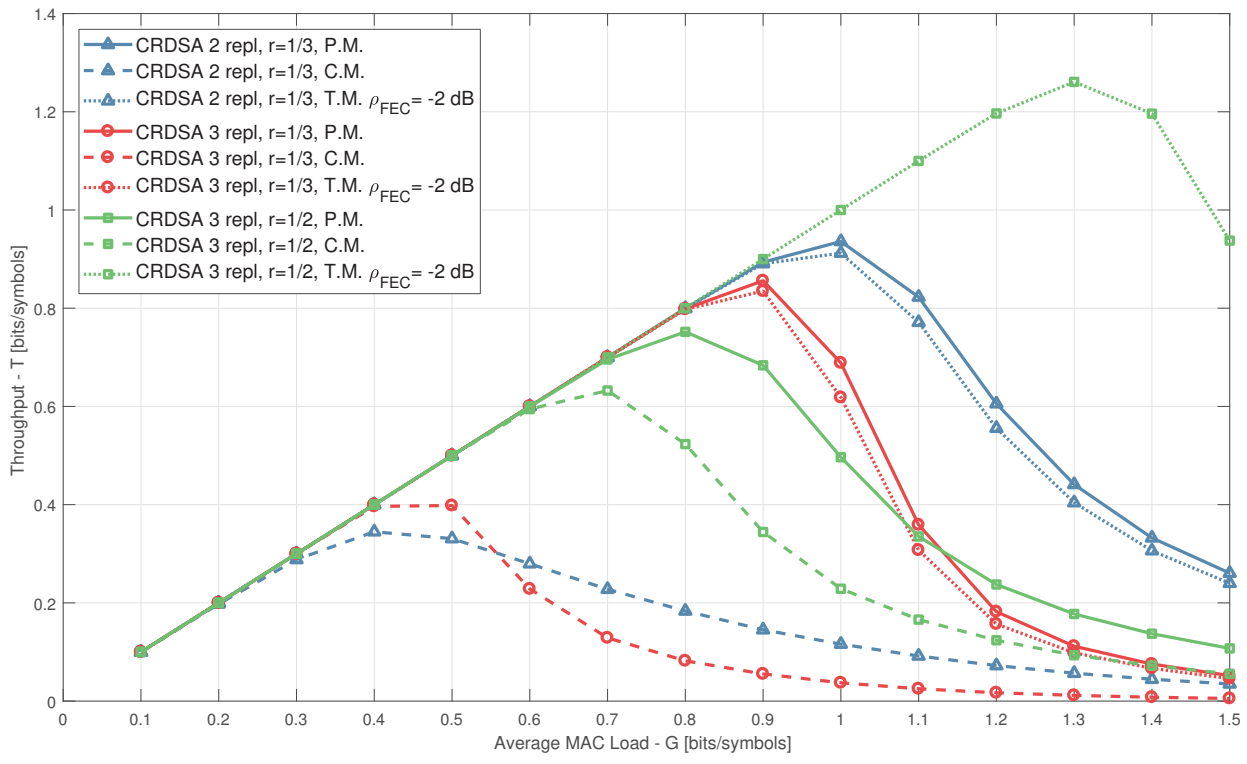

(a) Throughput

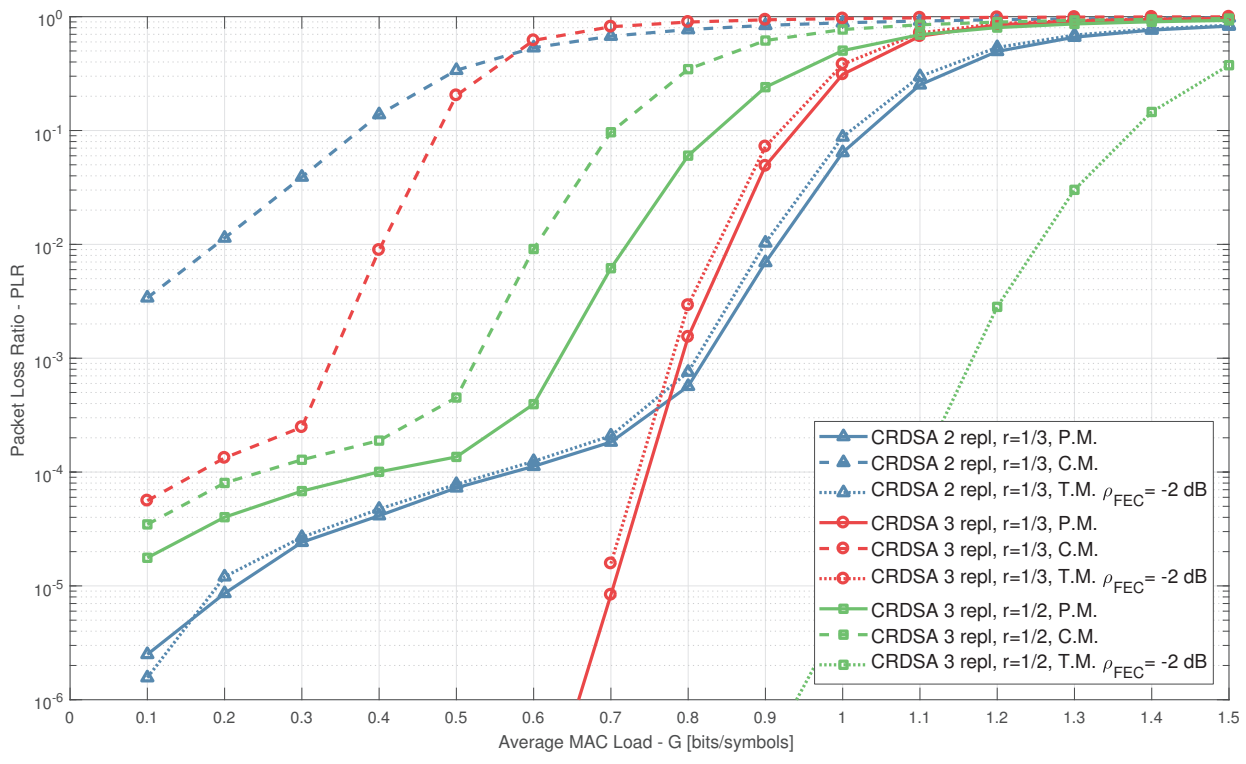

(b) PLR

Fig. 2. Comparison of simulated CRDSA throughput and PLR with Poisson traffic, as a function of the average load $G$ for 2 and 3 replicas; collision model (C.M.), threshold-based (T.M.) model with $\rho_{\text{FEC}} = -2$ dB and accurate PHY model (P.M.), $r = 1/3$ and $r = 1/2$ FEC, QPSK modulation, $N_{\text{slots}} = 128$, $N_{\text{iter}} = 20$, $E_s/N_0 = 10$ dB.

To further highlight the dependency on the system parameters of the match between threshold-based and accurate PHY models, results comparing these two models for values of $E_s/N_0 \in \{4, 6, 10\}$ dB are shown in Fig. 3 for CRDSA with 2 and 3 replicas using FEC rates 1/2 and 1/3. To reduce the number of lines in the plot, for each FEC rate, only the curve relative to the simplified model providing the closest match and the two most significant $E_s/N_0$ have been retained. It is important to note that the selected threshold value ($\rho_{\text{FEC}} = -2$ dB) always corresponds to resolution of cardinality-2 collisions for all the considered $E_s/N_0$ values, hence the omission of a specific $E_s/N_0$ value in the legend. For $r = 1/3$, results with $E_s/N_0 = 10$ dB are shown in Fig. 2 and almost equivalent to the ones with $E_s/N_0 = 6$ dB and are therefore not reported. We can observe that when $E_s/N_0 = 4$ dB, the difference between threshold-based and accurate PHY model is increasing. This is probably due to a higher probability of error in the case of cardinality-2 collision that is not correctly represented by the threshold-based model. On the contrary, for $r = 1/2$, when $E_s/N_0$ is reduced from 10 dB to 6 dB (results for $E_s/N_0 = 4$ dB were almost equivalent to the ones with $E_s/N_0 = 6$ dB), every collision becomes unresolvable and, differently from Fig. 2, the accurate PHY model results closely follows the collision model ones.

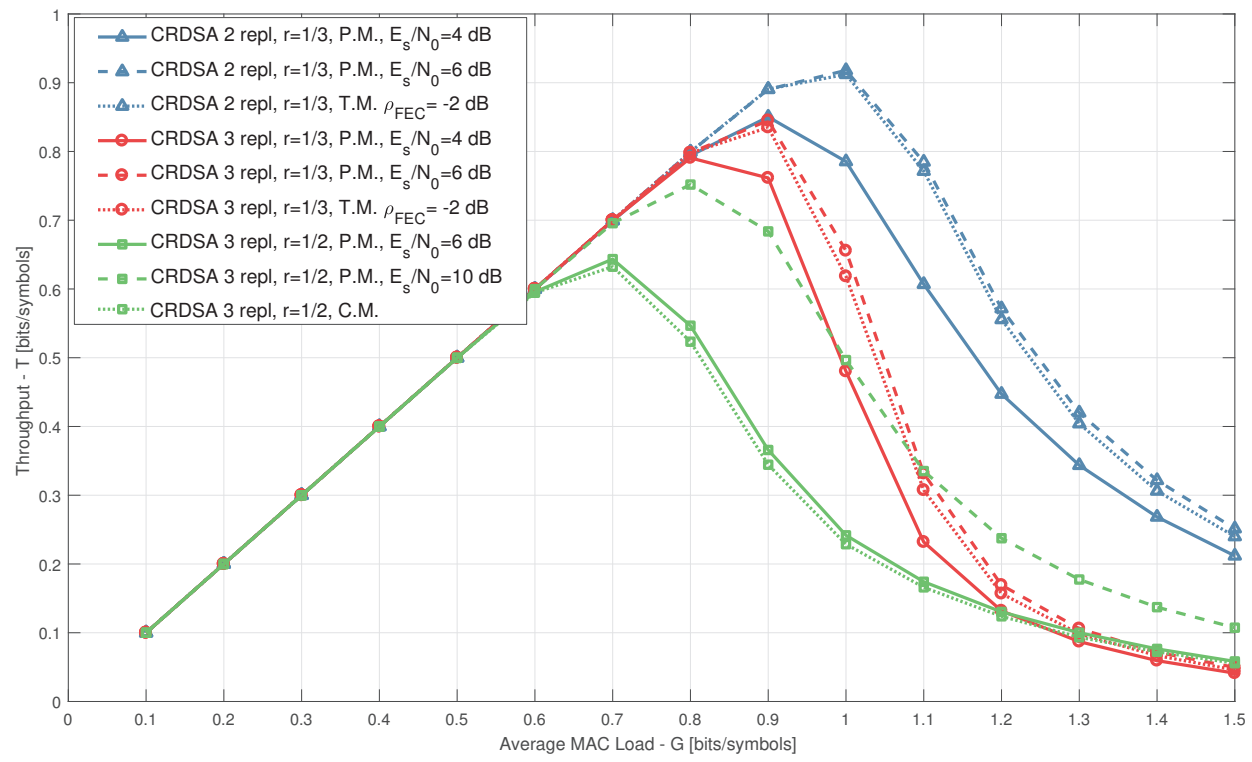

(a) Throughput

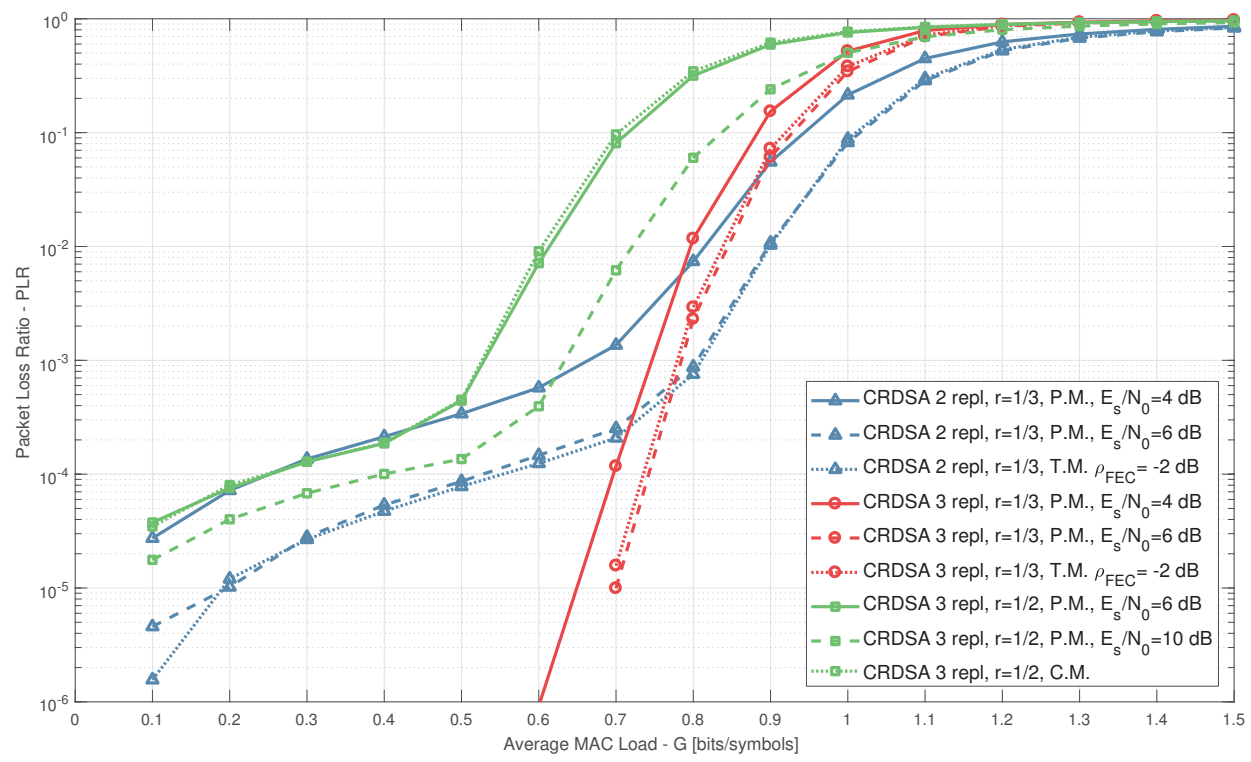

(b) PLR

Fig. 3. Comparison of simulated CRDSA throughput and PLR with Poisson traffic, as a function of the average load $G$ for 2 and 3 replicas; QPSK modulation, $N_{\text{slots}} = 128$, $N_{\text{iter}} = 20$. Results with $r = 1/3$ FEC are shown for accurate PHY model (P.M.) with $E_s/N_0 \in \{4, 6\}$ dB and compared against threshold-based model (T.M.) with $\rho_{\text{FEC}} = -2$ dB. Results with $r = 1/2$ FEC are shown for accurate PHY model with $E_s/N_0 \in \{6, 10\}$ dB and compared against collision model (C.M.).

The impact of non-equal, uniform in dB distribution of the received packets power on the threshold-based model with FEC rate $r = 1/3$ is shown in Fig. 4. As there is no analytical approach at hand to determine the right threshold value to be used, and in the power unbalanced case thresholds are not directly related to collision cardinalities, we resorted to the tedious trial and error approach to find the value of $\rho_{\text{FEC}}$ that best matches the characteristics of the accurate PHY model. Obviously, while $\rho_{\text{FEC}} = -2$ dB still provides the closest results to ones obtained via the accurate PHY model, its accuracy is diminished, compared to the scenario with equal powers of received packets shown in Fig. 4.

In conclusion, it can be stated that the performance results obtained with the collision channel model are very inaccurate in the considered scenarios. The threshold-based model can provide good prediction of CRDSA performance *only* if the correct threshold value is found for each specific scenario (e.g. FEC code rate 1/3, equal packet power level, $E_s/N_0 \geq 6$ dB). In other words, the threshold-based model has to be used with caution.

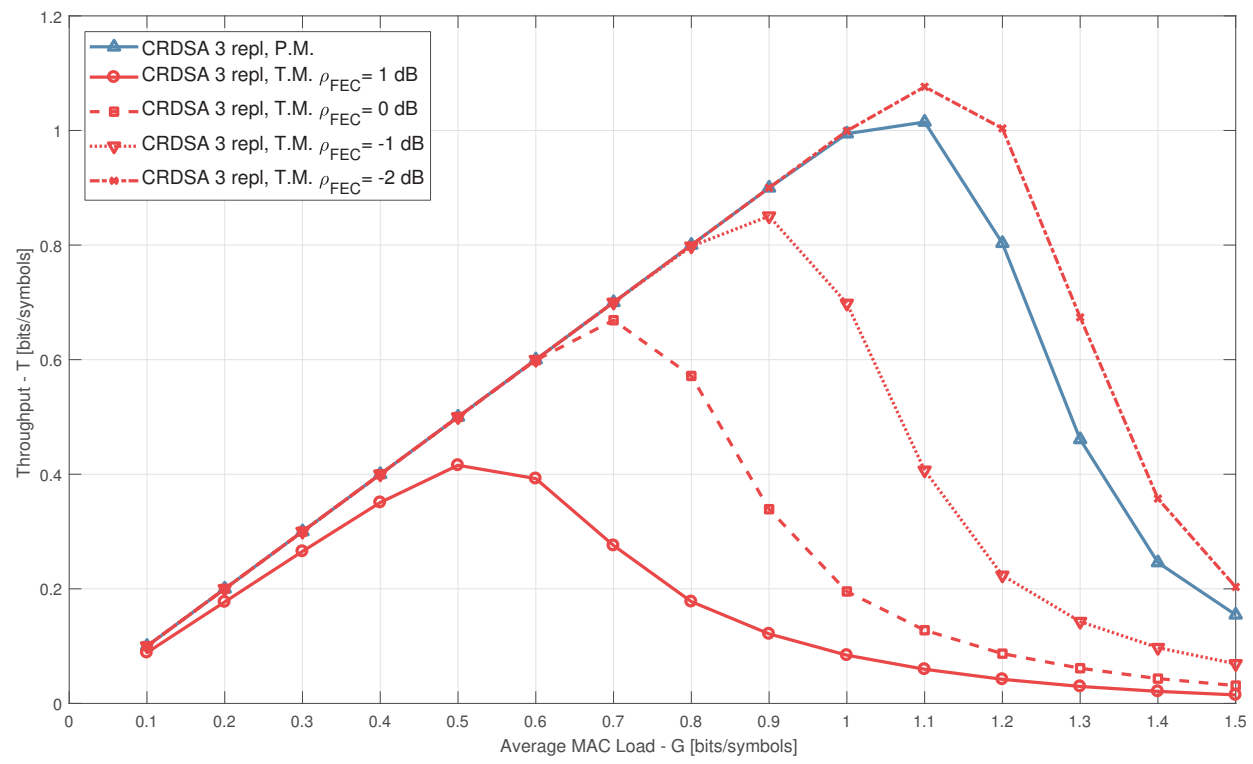

(a) Throughput

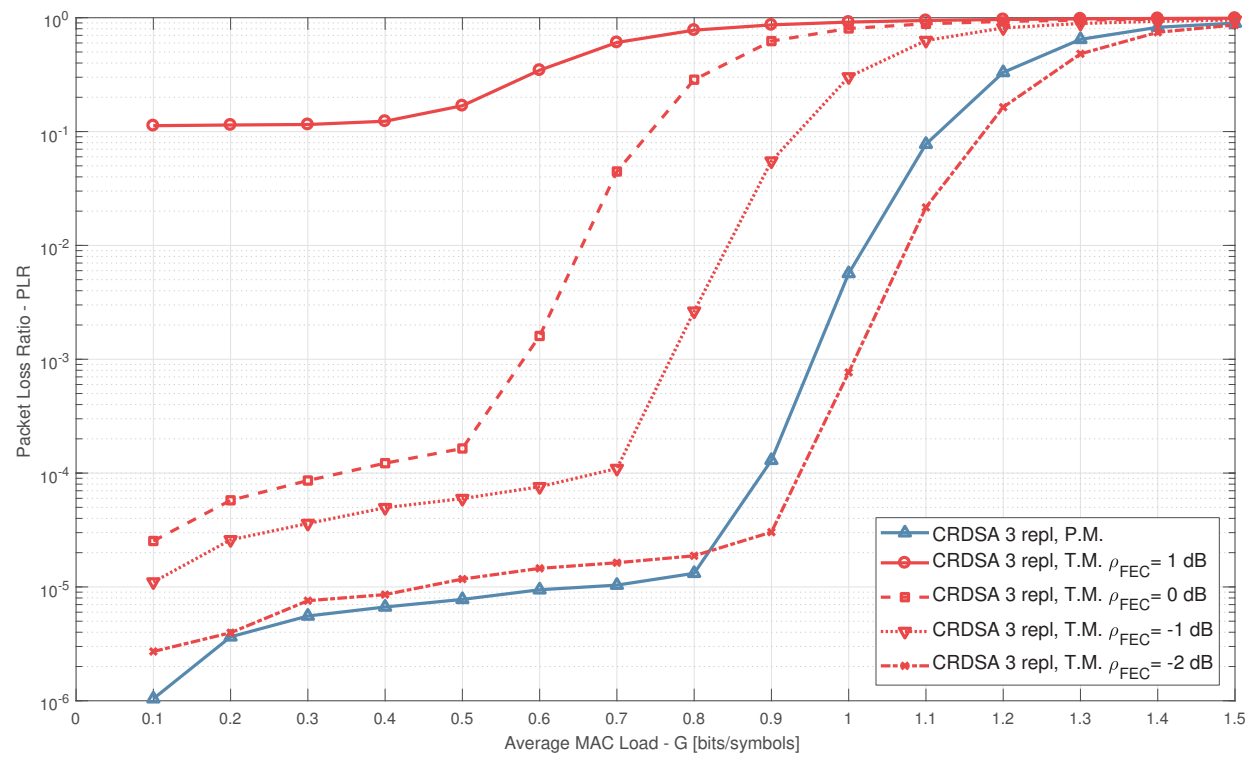

(b) PLR

Fig. 4. Comparison of simulated CRDSA throughput and PLR with Poisson traffic, as a function of the average load $G$ for 3 replicas; threshold-based model (T.M.) versus accurate PHY model (P.M.), $r = 1/3$ FEC, QPSK modulation, $N_{\text{slots}} = 128$, $N_{\text{iter}} = 20$, $E_b/N_0$ uniformly distributed between 2 and 9 dB.

## B. IRSA

We now introduce results obtained through our simulator using the IRSA scheme originally proposed in [9], with the following pmfs of the number of replicas transmitted by user:

- the scheme with pmf $\Gamma_3(x) = 0.5x^2 + 0.28x^3 + 0.22x^8$ [9], denoted by IRSA-1;[5]
- the scheme with $\Gamma_4(x) = 0.25x^2 + 0.6x^3 + 0.15x^8$ [9], denoted by IRSA-2;
- the scheme from with $\Gamma_5(x) = 0.87x^3 + 0.13x^8$ [21], denoted by IRSA-3.

We will focus on the variant with FEC of $r = 1/3$, as it outperforms the variant with $r = 1/2$ [7].

*1) Constant vs Poisson Traffic:* Similarly to CRDSA, the performance of IRSA for constant traffic model is better than for Poisson traffic model, for the same value of the average normalized load. Fig. 5 shows that, for Poisson traffic, the throughput at PLR $= 10^{-3}$ is about 13 % worse than for constant traffic. In the following, we use a more realistic Poisson traffic model.

[5]We use the standard polynomial notation, in which the power of a term denotes the number of replicas and the coefficient of the same term the probability of transmitting that many replicas.

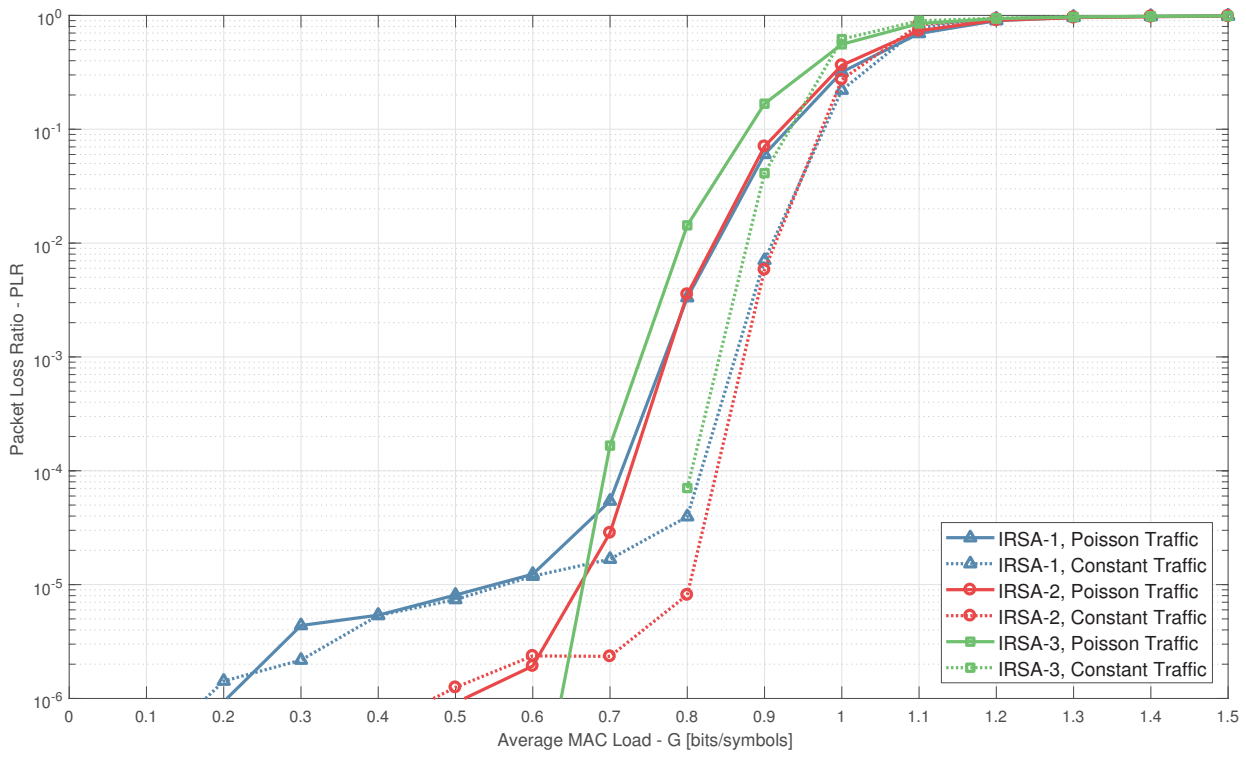

Fig. 5. Comparison of simulated IRSA-1/2/3 throughput and PLR with Poisson and constant traffic, as a function of the average load $G$; accurate PHY model, $r = 1/3$ FEC, QPSK modulation, $N_{\text{slots}} = 128$, $N_{\text{iter}} = 20$, $E_s/N_0 = 10$ dB.

*2) Simple Reception Models vs Accurate PHY Model:* As for CRDSA, we now investigate the difference between the accurate PHY model and the simplified models described in Sec. IV-C. The results for IRSA-2 with FEC code rate $1/2$ and $1/3$ are presented in Fig. 6; obviously, there is a trend similar to the results presented for CRDSA. For FEC $r = 1/3$ and $E_s/N_0 = 10$ dB, the collision model gives very pessimistic results, i.e., 260 % lower throughput than PHY model. Simulations of accurate PHY model with $r = 1/3$ and $E_s/N_0 = 10$ dB can be closely matched by the threshold-based model with $\rho_{\text{FEC}} = -2$ dB, but such an approximation could not be achieved for $r = 1/2$.

The results with IRSA-2 with FEC code rate $1/2$ and $1/3$ comparing accurate PHY and best-matching simplified model with various values of $E_s/N_0$ are presented in Fig. 7. Also in this case, we can see a dependency of the match between models on the value of $E_s/N_0$. This is in-line with results obtained for CRDSA and confirms the caution that has to be taken when using simplified reception models.

### C. CSA

We first validated the developed CSA simulator against the results reported in [10]. To do so, we recreated the same simulation conditions of the reference paper; namely constant traffic load (no Poisson) and the simple collision model. The results are not shown in this paper to limit the already significant number of figures, but we note that they closely match the ones reported in [10][6]. The details of the CSA scheme simulated, like the used codebooks and the probability of each code being selected, are specified in Table II and Eq. (28) in [10].

[6]It was found that throughput results reported in [10] were actually computed based on the Segment Loss Rate rather than on the PLR, implying that the successfully decoded segments of unsuccessfully decoded packets contribute to the throughput. In contrast, results reported here assume that the throughput corresponds only to the packets fully reconstructed by the demodulator, which is a more realistic approach.

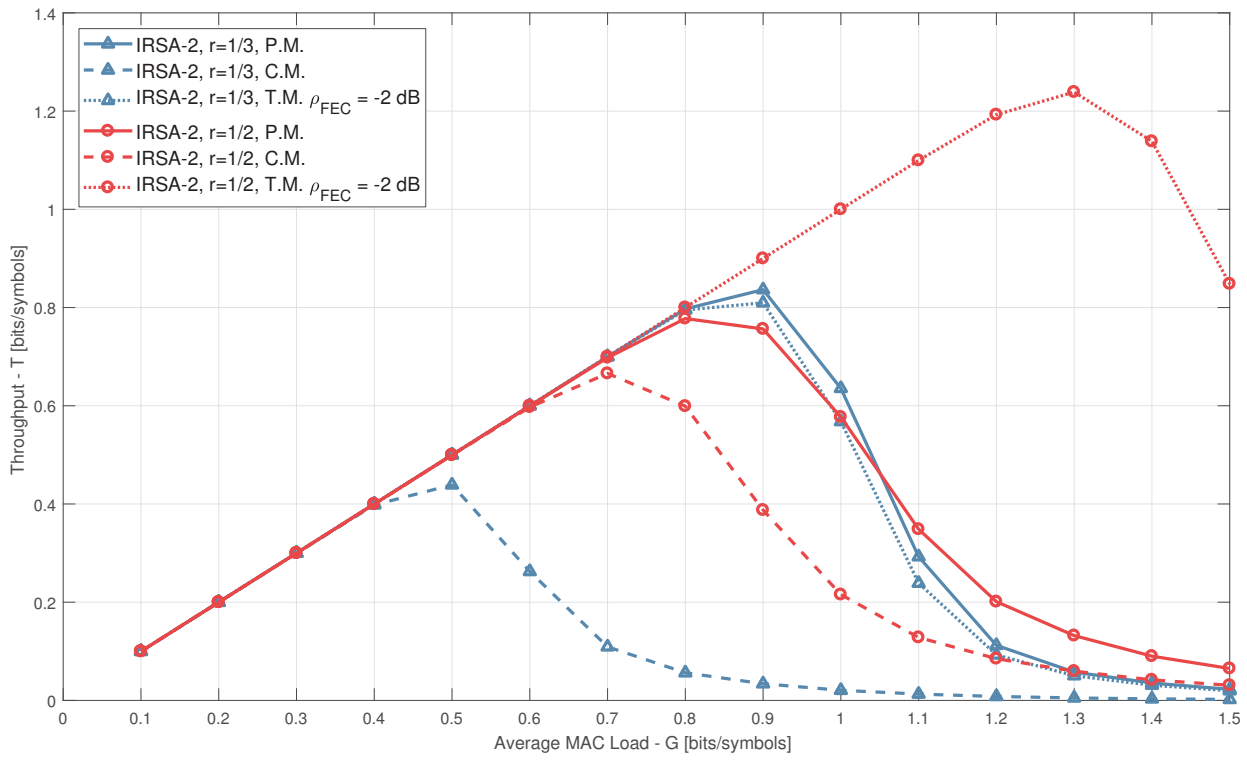

(a) Throughput

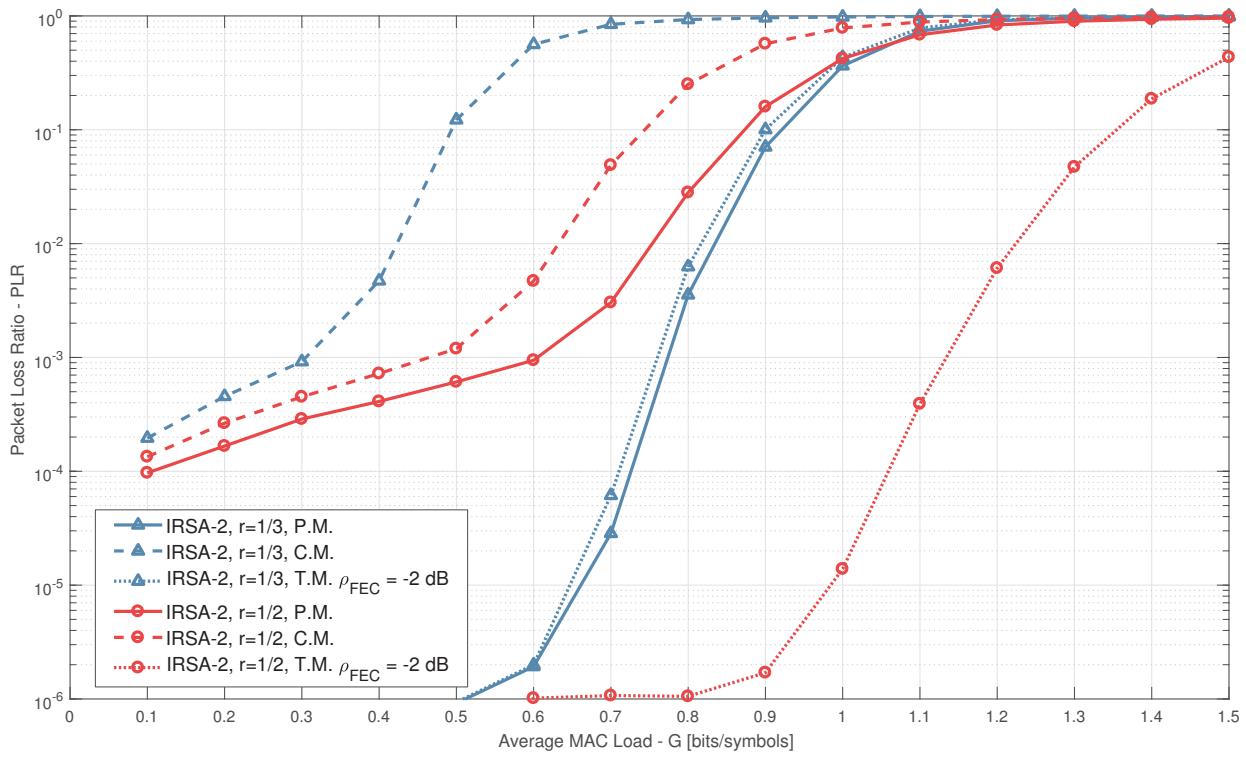

(b) PLR

Fig. 6. Comparison of simulated IRSA-2 throughput and PLR with Poisson traffic, as a function of the average load $G$; collision model (C.M.), threshold-based model (T.M.) with $\rho_{\text{FEC}} = -2$ dB and accurate PHY model (P.M.), $r = 1/2$ and $r = 1/3$ FEC, QPSK modulation, $N_{\text{slots}} = 128$, $N_{\text{iter}} = 20$, $E_s/N_0 = 10$ dB.

*1) Constant vs Poisson Traffic:* Fig. 8 shows that for Poisson distributed traffic exploiting the PHY model, the throughput at PLR $= 10^{-3}$ is about the same than with constant traffic for all three CSA configuration considered. Instead, for PLR $\geq 5 \cdot 10^{-3}$, similarly to the other schemes investigated Poisson traffic gives worst results. In the following, we use the more realistic Poisson traffic model.

*2) Simple Reception Models vs Accurate PHY Model:* As for CRDSA and IRSA, we now investigate the difference between the accurate PHY model and the simplified reception models described in Sec. IV-C, for the 3GPP turbo FEC with rate $r = 1/2$ [20] and QPSK modulation. The CSA results, presented in Fig. 9, are clearly pessimistic compared to the findings obtained by the accurate PHY model. To achieve meaningful PLR and throughput results, the need to use the accurate PHY model is evident. At a PLR $= 10^{-3}$ and for CSA with block-code rate $R = 1/3$, the collision model gives about 12 % lower throughput estimate than the accurate PHY model. Instead, the threshold model with $\rho_{\text{FEC}} = -2$ dB gives a throughput estimate 410 % better than the PHY's one.

The results for CSA with block-code rates $R \in \{1/3, 1/2, 3/5\}$, comparing collision and accurate PHY model at different values of $E_s/N_0$ are presented in Fig. 10. It is clear that the results obtained by collision and threshold models that match the ones obtained by using the accurate PHY model can only be achieved at $E_s/N_0 = 6$ dB.

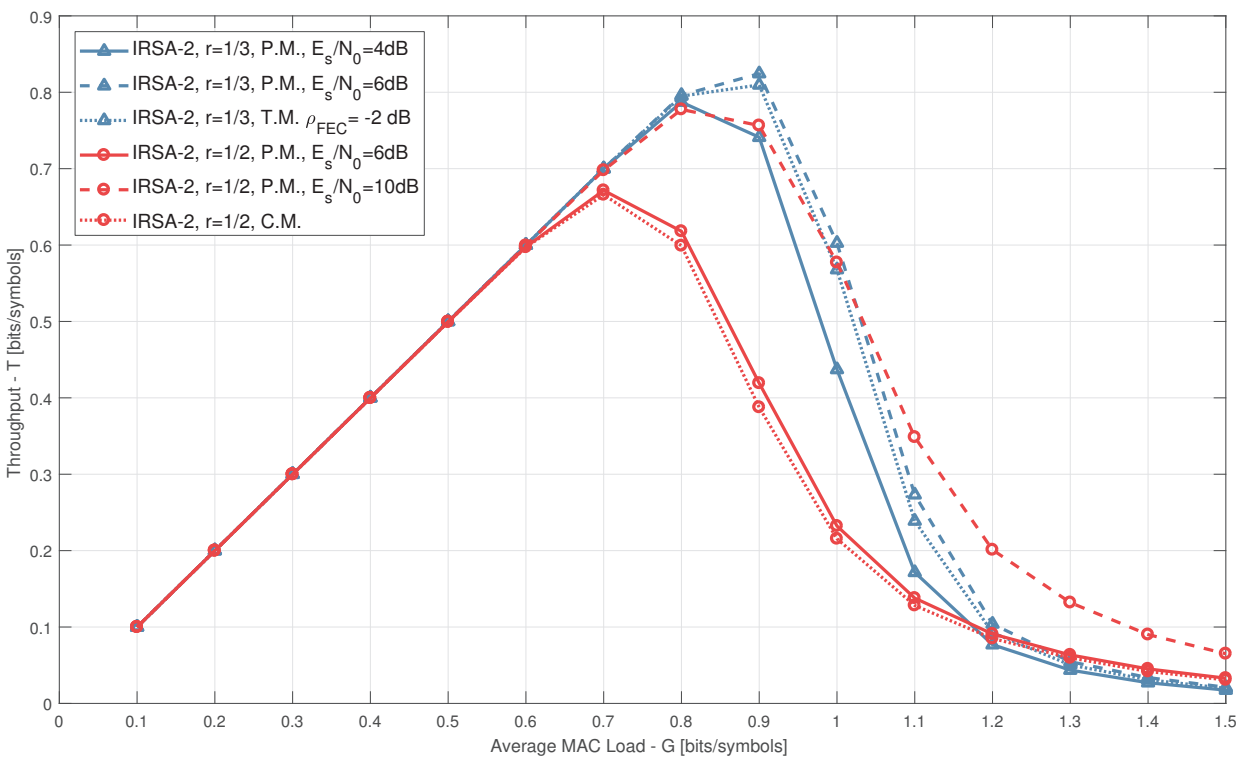

(a) Throughput

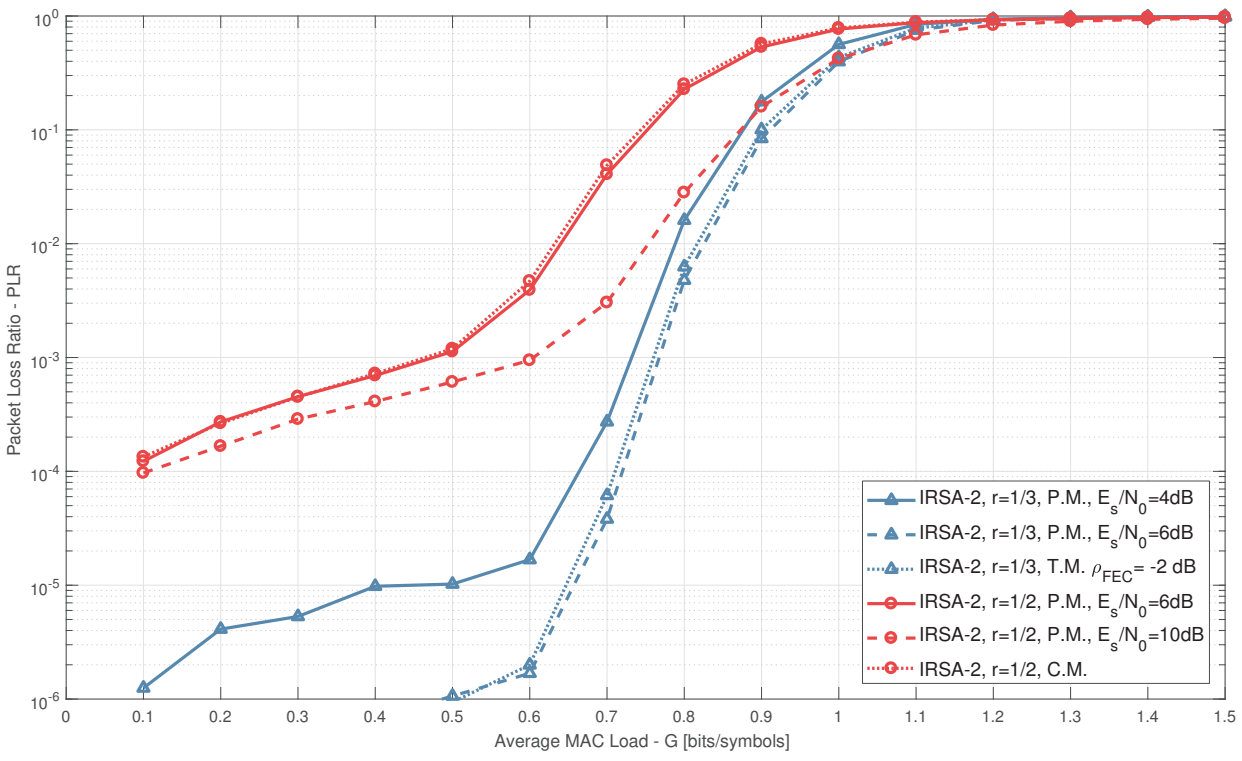

(b) PLR

Fig. 7. Comparison of simulated IRSA-2 throughput and PLR with Poisson traffic as a function of the average load $G$; QPSK modulation, $N_{\text{slots}} = 128$, $N_{\text{iter}} = 20$. Results with $r = 1/3$ FEC are shown for accurate PHY model (P.M) with $E_s/N_0 \in \{4, 6\}$ dB and compared against threshold-based model (T.M) with $\rho_{\text{FEC}} = -2$ dB. Results with $r = 1/2$ FEC are shown for accurate PHY model with $E_s/N_0 \in \{6, 10\}$ dB and compared against collision model (C.M).

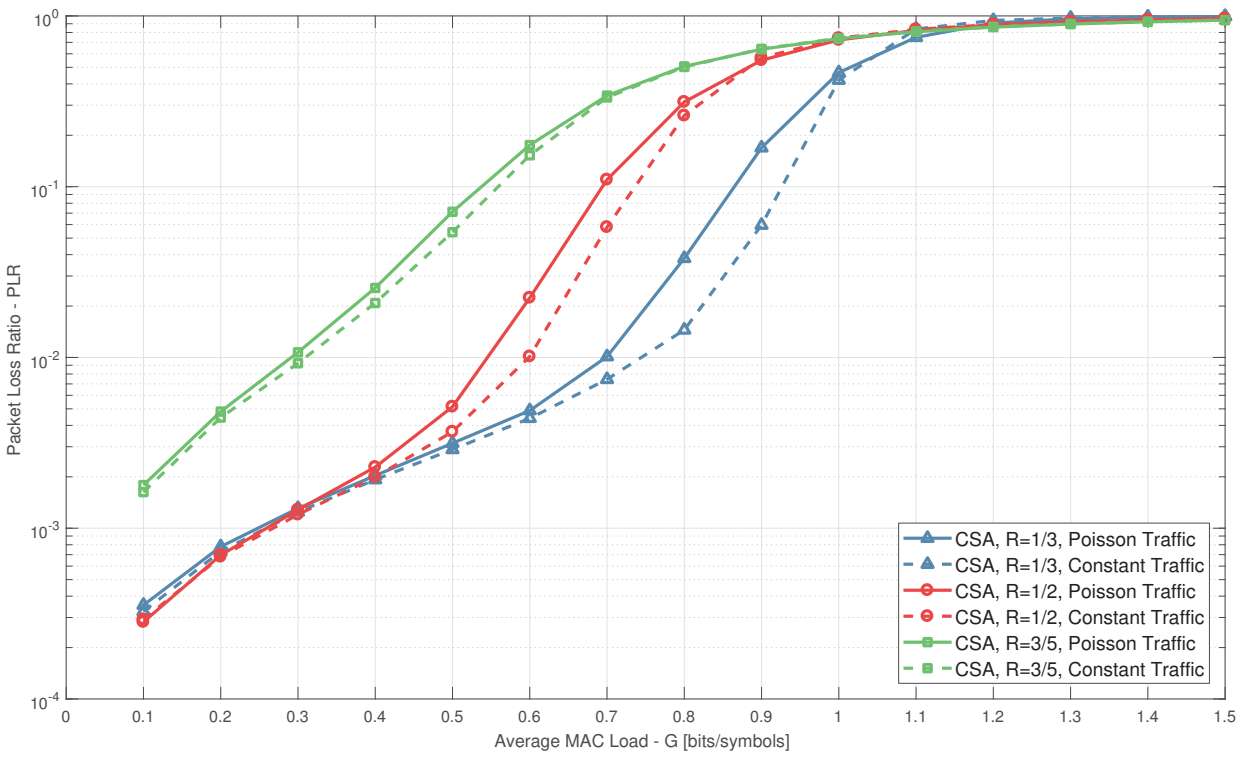

Fig. 8. Comparison of simulated CSA PLR with Poisson and constant traffic, as a function of the average load $G$; accurate PHY model, $r = 1/2$ FEC, CSA block-code rate $R \in \{1/3, 1/2, 3/5\}$, QPSK modulation, $N_{\text{slots}} = 128$, $N_{\text{iter}} = 20$, $E_s/N_0 = 10$ dB.

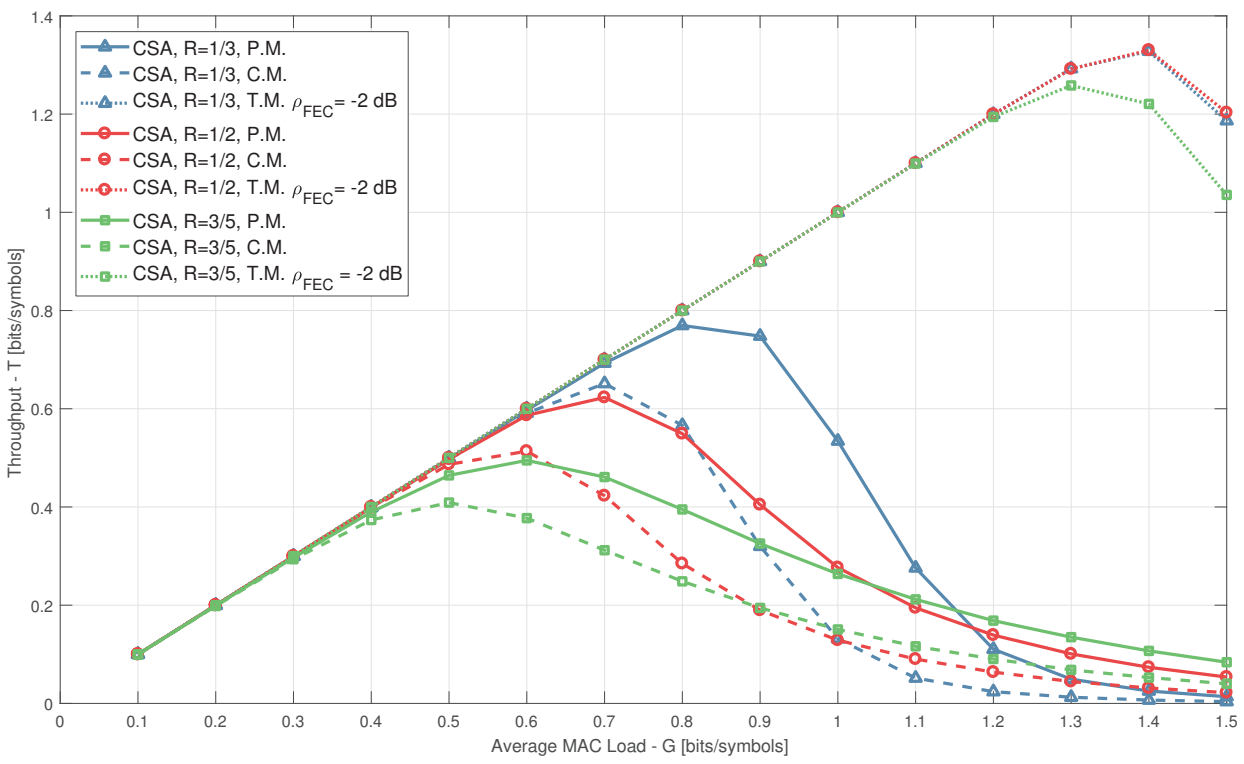

(a) Throughput

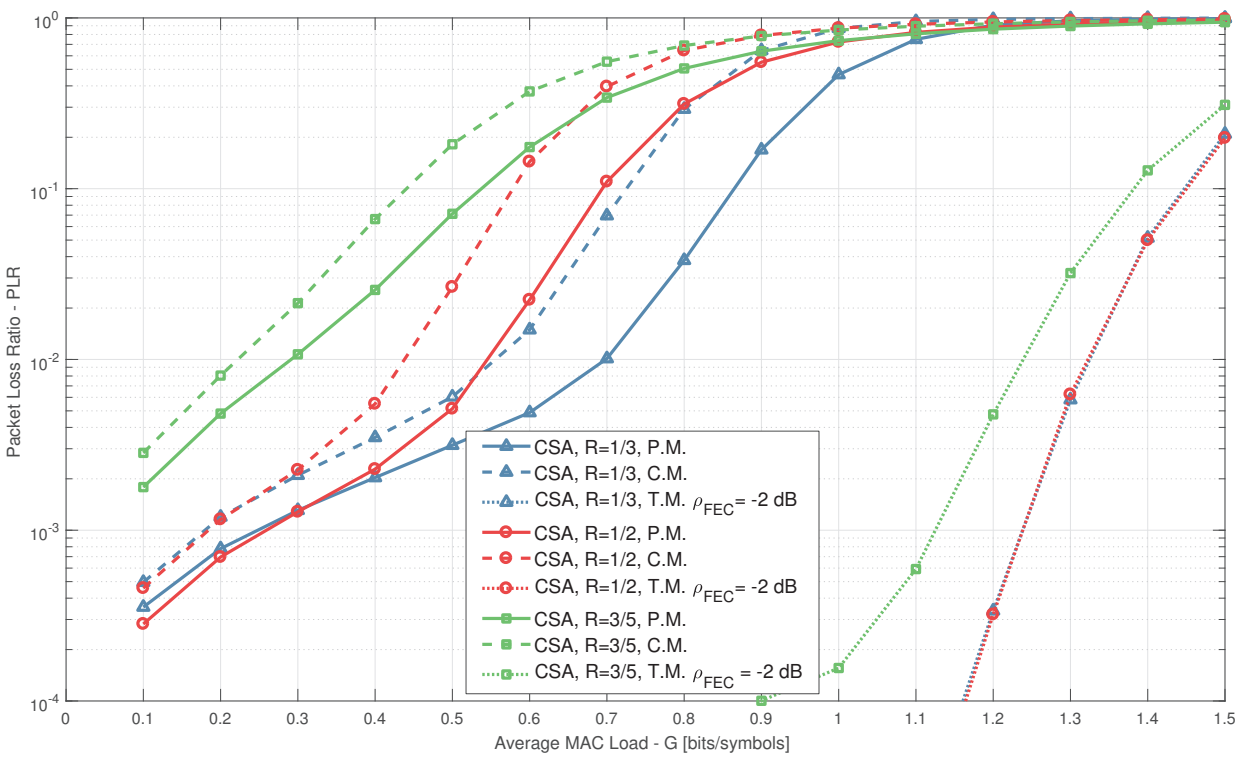

(b) PLR

Fig. 9. Comparison of simulated CSA throughput and PLR with Poisson traffic, as a function of the average load $G$; collision model (C.M.), threshold-based model (T.M.) with $\rho_{\text{FEC}} = -2$ dB and accurate PHY model (P.M.), $r = 1/2$ FEC, CSA block-code rate $R \in \{1/3, 1/2, 3/5\}$, QPSK modulation, $N_{\text{slots}} = 128$, $N_{\text{iter}} = 20$, $E_s/N_0 = 10$ dB.

### D. E-SSA

*1) Constant vs Poisson Traffic:* Fig. 11 shows the simulated E-SSA throughput and PLR with Poisson and constant traffic as a function of the average load $G$ with the accurate PHY model. It is apparent that, thanks to the large level of E-SSA traffic aggregation, there is a negligible difference between the Poisson and the constant traffic models. However, Fig. 11 also shows that at PLR $= 10^{-4}$ there is a slight PLR performance improvement when using a constant traffic assumption.

*2) Threshold-Based Reception Model vs Accurate PHY Model:* Fig. 12 compares the simulated E-SSA throughput and PLR with Poisson traffic as a function of the average load $G$, threshold-based and accurate PHY models. All packets are assumed to be received at the same power and an AWGN channel with $E_s/N_0 = 6$ dB has been considered. For the threshold-based model, different threshold values $\rho_{\text{FEC}}$ have been assumed. It is apparent that the results significantly depend on the selected value of $\rho_{\text{FEC}}$. The closest result is obtained for $\rho_{\text{FEC}} = -6$ dB and even better matching appears for $\rho_{\text{FEC}}$ between $-5$ and $-6$ dB, which is quite different from CRDSA/IRSA best matching threshold. This is due to the multi-packet reception that is consequence of spreading. For completeness, the SEA results from [11] (Sect. III.A) are also compared to the more accurate PHY layer Monte Carlo simulations obtained following the approach outlined in Sect. III.D of [11]. Fig. 13 shows that, when the optimum randomization of the received packet power (uniform in dB) derived in [16] is introduced, the results obtained with the threshold-based model results are even less accurate. Not only the threshold-based model is showing a very optimistic maximum load for PLR $= 10^{-3}$, but also the PLR floor present in the PHY model is absent. The origin of the PLR floor with the randomized packets' power is explained in [16]. This example shows that the threshold-based model can not be used in practice, since the threshold value approximating the real system behavior depends on many system parameters, including the packets power distribution. Summarizing, we have experimentally shown that finding a threshold value that will provide a good match between the threshold-based and the accurate PHY models in the equi-powered packets' case is laborious and does not apply to the case of randomized distribution of packets' powers.

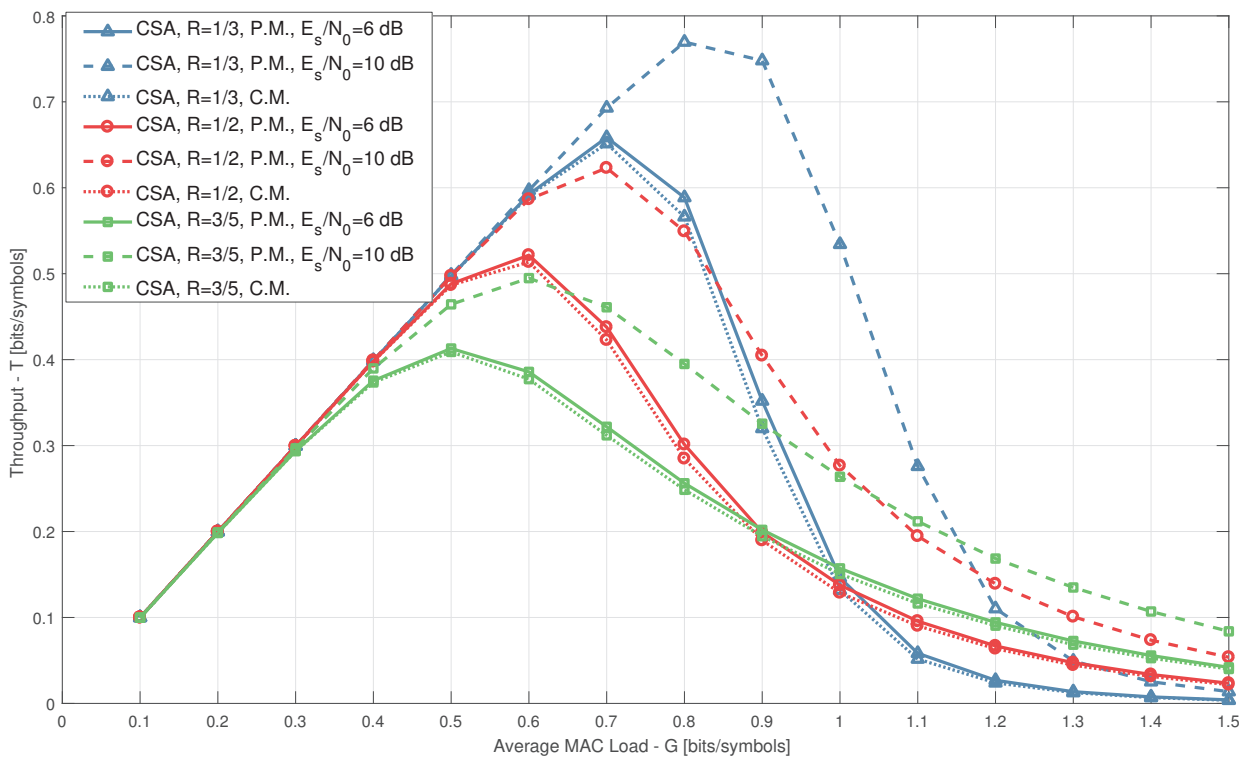

(a) Throughput

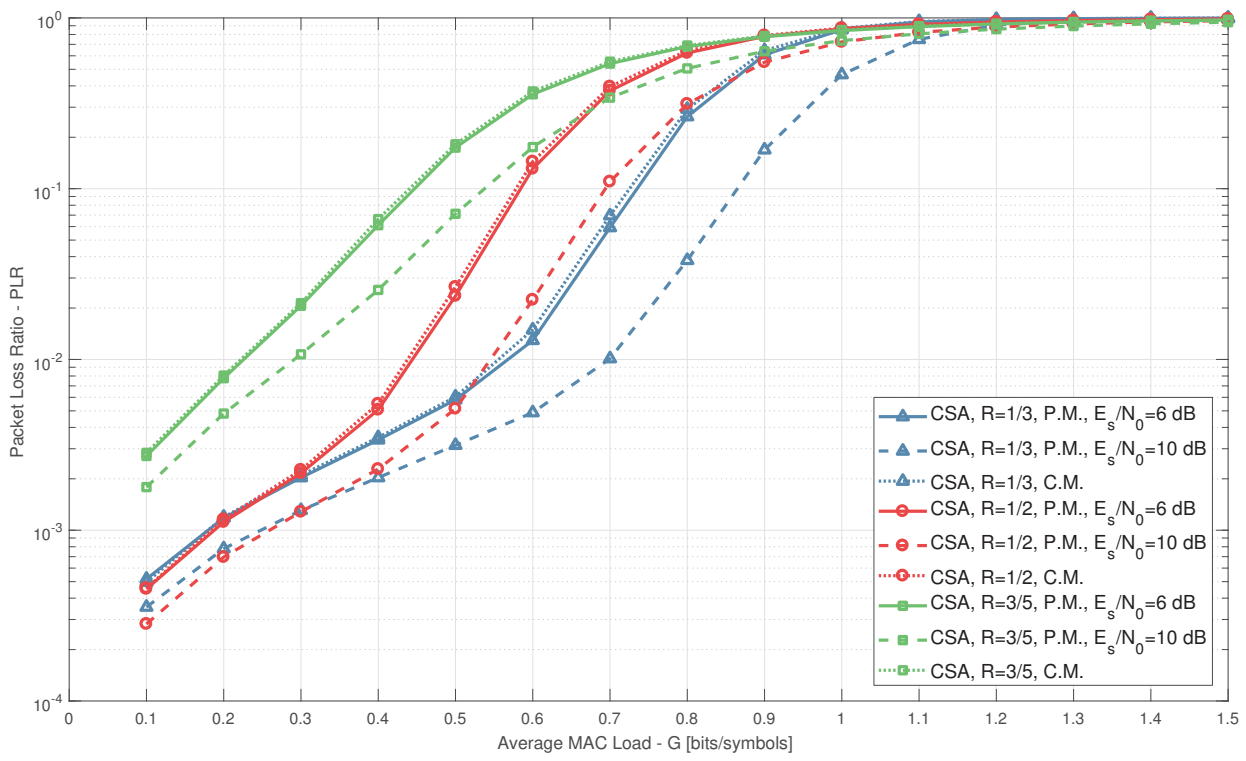

(b) PLR

Fig. 10. Comparison of simulated CSA throughput and PLR with Poisson traffic, as a function of the average load $G$; collision model (C.M.) and accurate PHY model (P.M.), $r = 1/2$ FEC, CSA block-code rate $R \in \{1/3, 1/2, 3/5\}$, QPSK modulation, $N_{\text{slots}} = 128$, $N_{\text{iter}} = 20$, $E_s/N_0 = 6,\ 10$ dB.

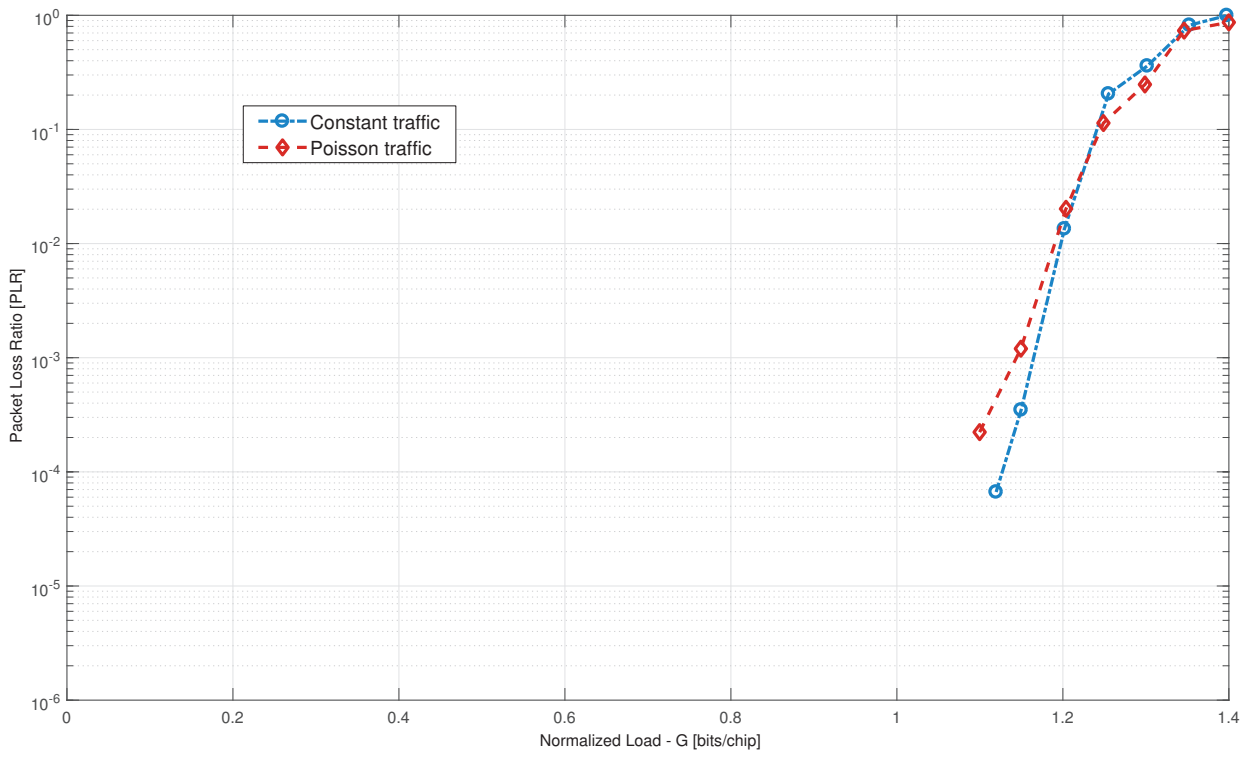

Fig. 11. Comparison of simulated E-SSA PLR with Poisson and constant traffic, as a function of the average load $G$; accurate PHY model (P.M.), $r = 1/3$ FEC, BPSK modulation, SF $= 64$, $N_{\text{iter}} = 10$, $E_s/N_0 = 6$ dB.

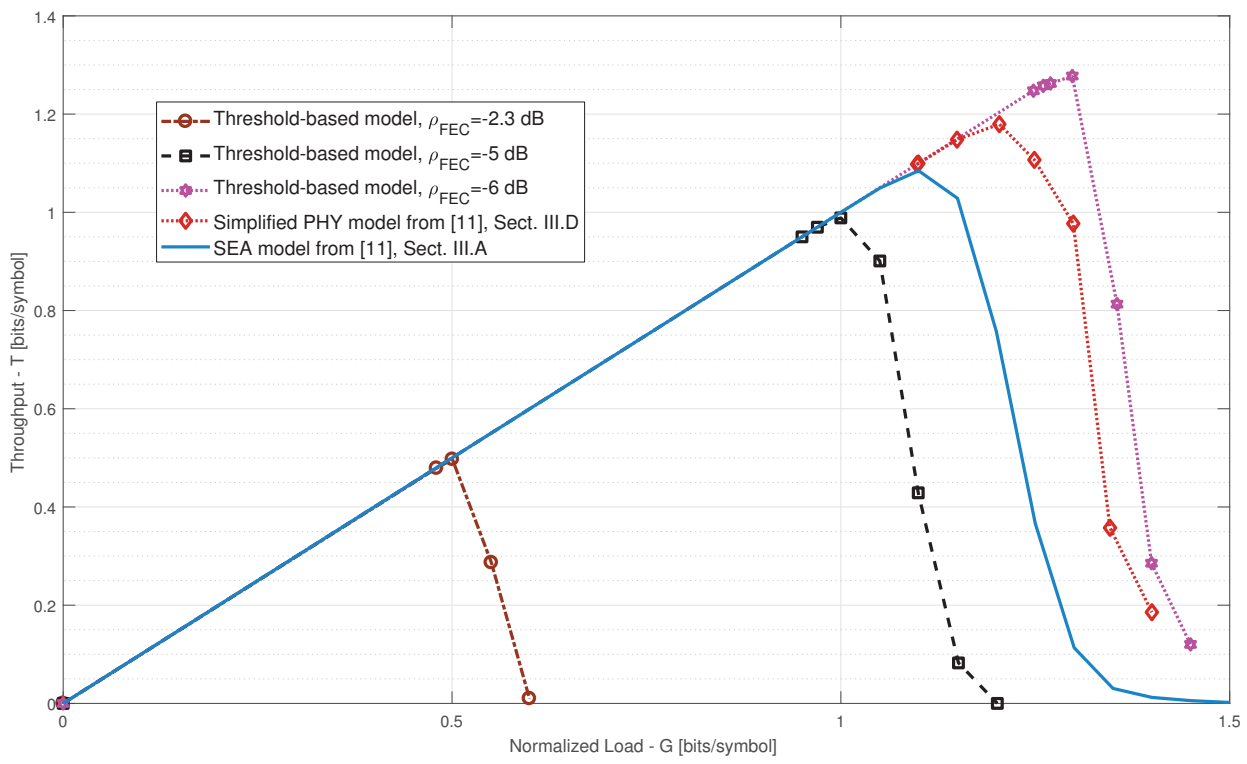

(a) Throughput

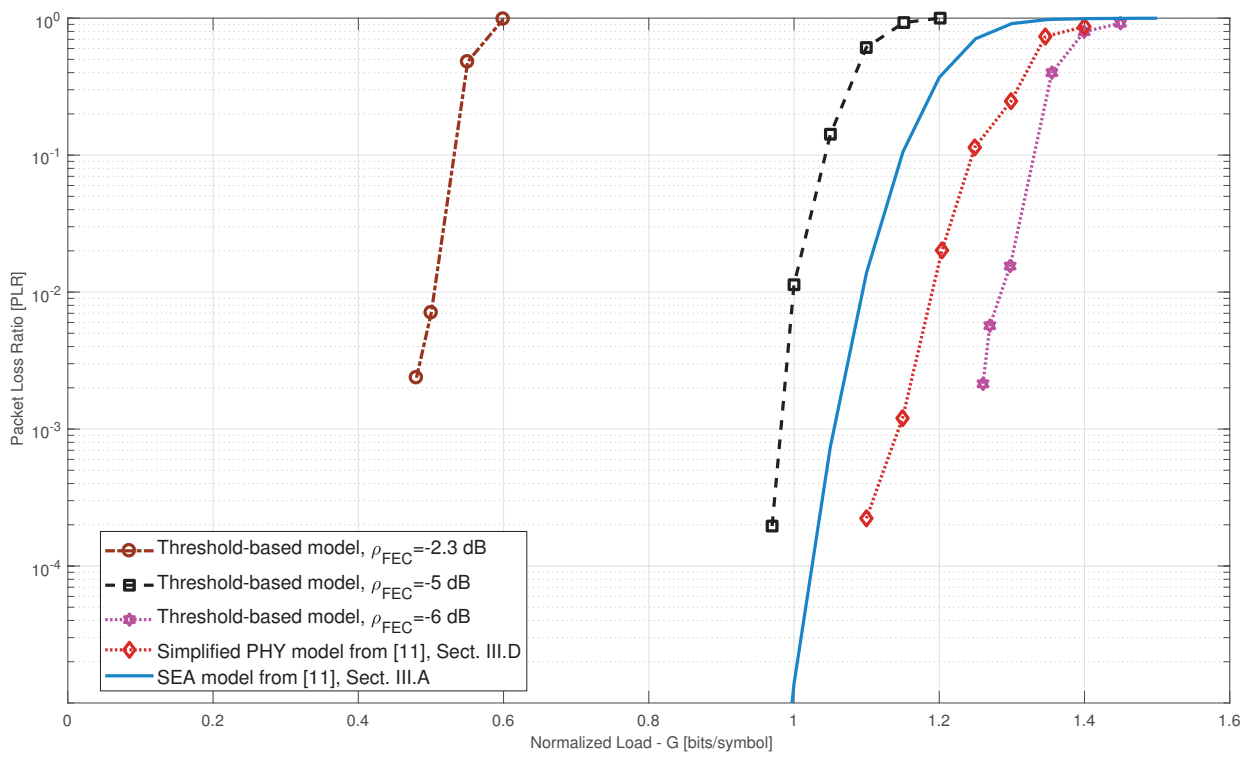

(b) PLR

Fig. 12. Comparison of simulated simplified physical model (PHY) and semi-analytical (SEA) E-SSA throughput and PLR with Poisson traffic, as a function of the average load $G$; threshold-based and accurate PHY models, $r = 1/3$ FEC, BPSK modulation, SF $= 64$, $N_{\text{iter}} = 10$, $E_s/N_0 = 6$ dB.

## *E. RA Schemes comparison*

Here we provide a fair comparison of the performance parameters of the schemes under investigation, following Sec. III methodology. The performance comparison exercise has been done for the most accurate model available i.e., the accurate PHY model for CRDSA, IRSA, CSA and the Monte Carlo simulation for E-SSA. To limit the number of cases to be simulated for each slotted RA scheme, we selected the best performing configuration in terms of code rate and number of replicas. For fairness, all schemes share the same 3GPP turbo FEC scheme previously introduced. Following Sec. III, the E-SSA spreading factor is half of the number of slots per frame (i.e., SF $= 64$ while $N_{\text{slots}} = 128$). To keep the same average packet power per frame as for Sec. III-D, we used results from slotted RA schemes simulations with the same $E_s/N_0 = 6$ dB adopted for E-SSA. The corresponding simulation results are summarized in Table I. It appears that E-SSA is outperforming all other schemes both in terms of throughput and energy efficiency at target PLR $= 10^{-3}$. The runner-up scheme is CRDSA with 2 replicas and $r = 1/3$. But its throughput and energy efficiency is respectively 29 and 100 % lower than E-SSA. In terms of normalized throughput, CSA performs worse than CRDSA, while its energy efficiency is 50 % better than CRDSA. We believe that both IRSA and CSA normalized throughput performance are penalized by the fact that their design was optimized using the inaccurate collision model.

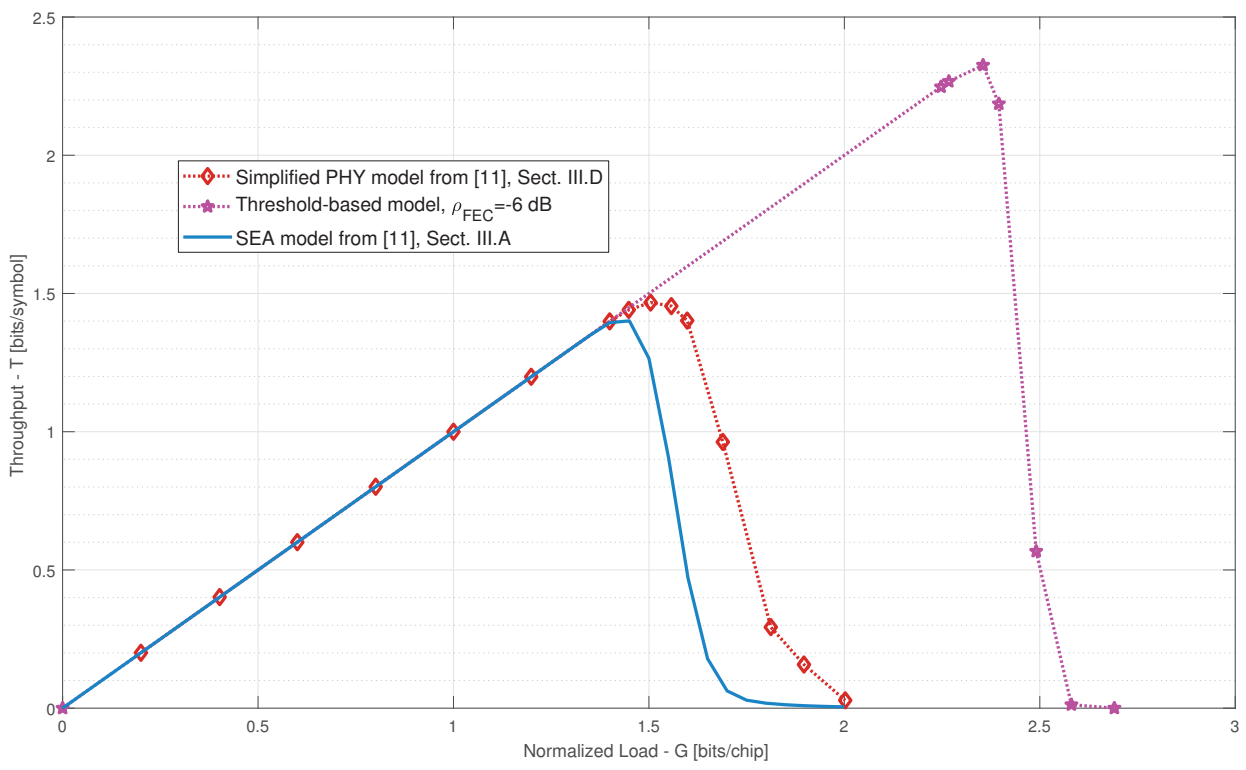

(a) Throughput

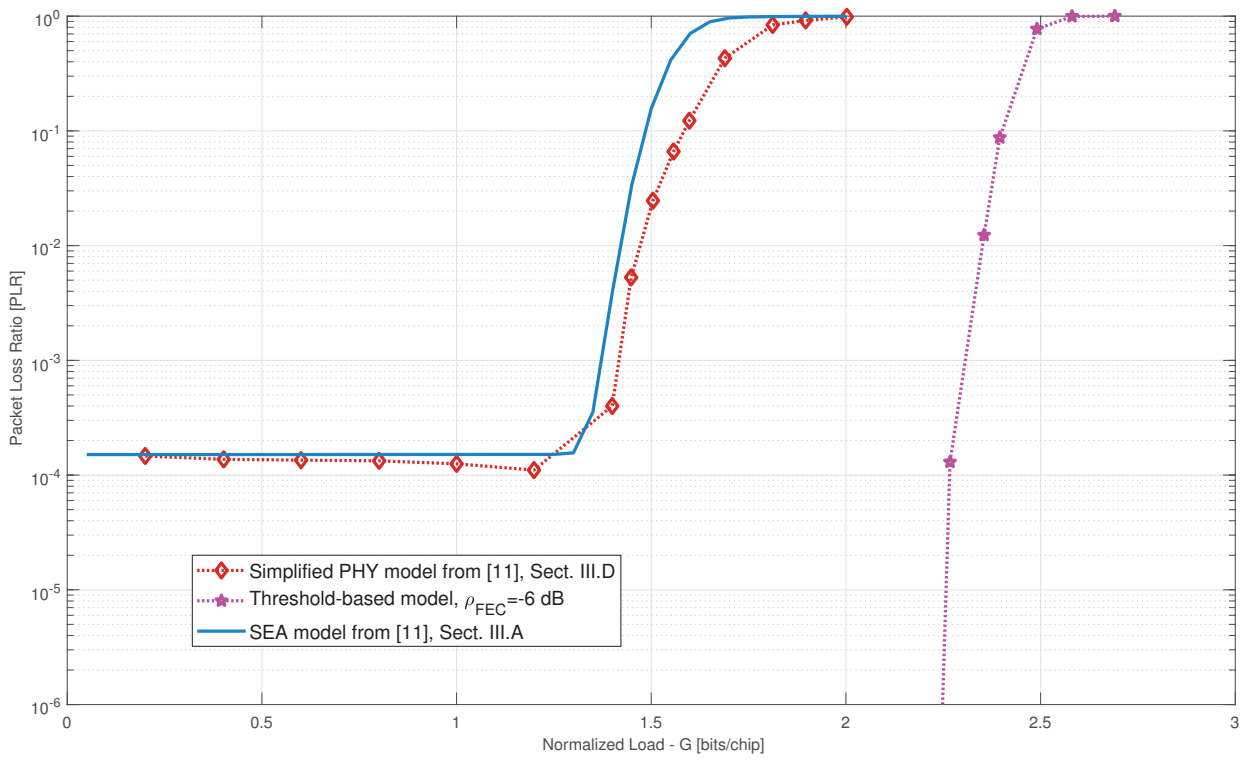

(b) PLR

Fig. 13. Comparison of simulated simplified physical model (PHY) and semi-analytical (SEA) E-SSA throughput and PLR with Poisson traffic, as a function of the average load $G$, threshold-based model and accurate PHY model, $r = 1/3$ FEC, BPSK modulation, SF = 64, $N_{\mathrm{iter}} = 10$, with uniform distribution in dB in the range [2 - 9] dB.

## VI. Summary and Recommendations

In this paper, we thoroughly reviewed different methodologies widely adopted in literature for assessing the RA performance with the following main findings:

- Poisson traffic has an impact on CRDSA, IRSA and CSA schemes, i.e., there is up 10 % throughput reduction compared to a constant traffic model with the same (average) load in the examined scenarios. For E-SSA, as expected, the impact is instead negligible, due to the higher level of traffic aggregation at physical level.
- The collision channel model provides very unreliable results when physical layer coding is used in CRDSA, IRSA and CSA schemes. This simple model is not usable for E-SSA.
- The threshold-based model should be carefully optimized by finding a threshold value that matches accurate simulation results. The value of the threshold depends on the RA scheme, FEC coding rate and the distribution (i.e., unbalance) of the received power of the packets. In this respect, the threshold-based reception model seems to come short of proper modeling of CRDSA, IRSA and CSA with some specific code rates and signal-to-noise ratios where the packets have the same power at the point of reception.
- The semi-analytical model reported in [7] provides accurate results for slotted RA schemes such as CRDSA, IRSA and CSA with a much faster processing time compared to Monte Carlo simulations including the physical layer.
- For E-SSA with given physical layer parameters, finding the matching threshold model value is very laborious and results will become inaccurate if the incoming packets have different power levels.
- The E-SSA semi-analytic model described in [11] provides fairly accurate results for both equi-powered and randomly distributed powers of incoming packets.
- In general, the optimal threshold value for the threshold-based reception model has to be heuristically derived by trial and error method. This provides limited advantage compared to the existing semi-analytical models that include accurate model of the physical layer.
- The pmfs for IRSA and CSA that were optimized under the assumption of the collision channel model result in performance that is inferior to the performance of (simpler) CRDSA and E-SSA.
- We also derived a general formulation of the RA schemes energy efficiency. It was found that the best performing scheme is E-SSA, being 2 times more energy efficient than CRDSA and CSA, and 3.5 times more than IRSA. Further, E-SSA is also 29 % and 288 % more efficient in terms of normalized throughput than CRDSA and CSA, respectively.

Finally, we turn to an open question of optimizing IRSA and CSA schemes under the accurate physical layer model of satellite networking. A potential and general way to do this is to use the analytical optimization tool presented in [9] and incorporate into it the probabilities of decoding a packet from a collision of a given cardinality, cf. Appendix A of [9]. These probabilities can be derived using the accurate physical layer simulator. Another potential approach, which may be suitable for scenarios that can be accurately modeled using the threshold-based reception, is to (i) derive the probability distribution function of SNIR of a packet involved in a collision of a given cardinality, (ii) calculate the probability of decoding the packet using the optimized value of the threshold, and (iii) plug-in these probabilities in the optimization tool presented in [9]. This approach was employed in [5], [18], for the Rayleigh block fading channel, showing that such optimization grants significant increase in the throughput performance when compared to the variants of the scheme optimized for the collision channel model. Investigation of the approach used in [5], [18] for the optimization of IRSA and CSA for satellite networking is in focus of our on-going work.

| RA Scheme | $G_p$ | $N_{\mathrm{slots}}$ | $\overline{N}_{\mathrm{rep}}$ | Normalized Throughput $T$ [bits/symbol or bits/chip] | Energy Efficiency $\psi_e^n$ |
|---|---|---|---|---|---|
| CRDSA $r = 1/3$ | 3/2 | 128 | 2 | 0.81 | 2.60 $\cdot 10^{-3}$ |
| IRSA-2 $r = 1/3$ | 3/2 | 128 | 3.5 | 0.77 | 1.49 $\cdot 10^{-3}$ |
| CSA $r = 1/2$, $R = 1/2$ | 1 | 128 | 2 | 0.19 | 3.9 $\cdot 10^{-3}$ |
| E-SSA $r = 1/3$ | 3×64 | 1 | 1 | 1.15 | 5.21 $\cdot 10^{-3}$ |

TABLE I
ESTIMATED THROUGHPUT AND ENERGY EFFICIENCY USING THE PHY MODEL FOR THE VARIOUS RA SCHEMES FOR A TARGET PER$= 10^{-3}$; $E_s/N_0 = 6$ dB.

## ACKNOWLEDGEMENT

The authors would like to thank Prof. Petar Popovski for his helpful comments and insights during the preparation of this work and Dr. Enrico Paolini and Dr. Gianluigi Liva for their support in validating the CSA simulator.

## APPENDIX

### A. Slot and Frame Traffic-Distribution

To be able to adequately model the collisions that take place in the slots of a frame, we first model the distribution of the number of colliding packets in every slot. We start from the case in which the number of users in the slot is fixed and then move to the case when this number is Poisson distributed.

*1) Distribution of Collision Cardinality for a Fixed Number of Users:* Suppose a frame composed of $N_{\mathrm{slots}}$ slots in which $L$ users are transmitting packets with an average number of replicas per user equal to $\overline{N}_{\mathrm{rep}}$. As stated in [9], [10], we can express the probability $p$ of a generic user selecting a given slot for one of its replicas as

$$p = \frac{\overline{N}_{\mathrm{rep}}}{N_{\mathrm{slots}}}. \quad (13)$$

As every user chooses the slots independently, the number of packets colliding in the generic slot $s$ (i.e., the collision cardinality of $s$) is a random variable denoted as $I$, which follows a binomial distribution with parameters $L$ and $p$

$$I \in \mathcal{B}(L,p) \to Pr\{I = i\} = \binom{L}{i} p^i (1-p)^{L-i}. \quad (14)$$

*2) Distribution of Collision Cardinality for a Poisson Number of Users:* Consider now the same scenario, with the difference that the number of users active in the frame $L$ is not fixed anymore, but follows a Poisson distribution, described by the following equation

$$L \in \mathcal{P}(\lambda) \to Pr\{L = l\} = \frac{\lambda^l}{l!} e^{-\lambda}. \quad (15)$$

The distribution of $I$ is now Binomial with a number of trials that is Poisson distributed and not fixed:

$$Pr\{I = i\} = \sum_{l=i}^{\infty} \binom{l}{i} p^i (1-p)^{(l-i)} \frac{\lambda^l}{l!} e^{-\lambda}. \quad (16)$$

By expanding the binomial coefficient, we obtain

$$\sum_{l=i}^{\infty} \binom{l}{i} p^i (1-p)^{(l-i)} \frac{\lambda^l}{l!} e^{-\lambda} = \frac{p^i e^{-\lambda}}{(1-p)^i i!} \sum_{l=i}^{\infty} \frac{[\lambda(1-p)]^l}{(l-i)!}. \quad (17)$$

Applying the change of variables $m = l - i$ and $q = 1 - p$, we get

$$\frac{p^i e^{-\lambda}}{q^i i!} \sum_{l=i}^{\infty} \frac{(\lambda q)^l}{(l-i)!} = \frac{p^i e^{-\lambda}}{q^i i!} \sum_{m=0}^{\infty} \frac{(\lambda q)^{(m+i)}}{m!}. \quad (18)$$

Recalling that $\sum_{l=0}^{\infty} \frac{z^l}{l!} = e^z$, we can further simplify the previous expression as

$$\frac{p^i e^{-\lambda}}{q^i i!} \sum_{m=0}^{\infty} \frac{(\lambda q)^{(m+i)}}{m!} = \frac{p^i e^{-\lambda}}{q^i i!} (\lambda q)^i e^{\lambda q} = \frac{(\lambda p)^i e^{-(\lambda p)}}{i!}. \quad (19)$$

We can see from equation (19) that the resulting distribution is also Poisson with mean value $\lambda \cdot p$.

PLACE PHOTO HERE

**Alberto Mengali** was born in Pisa in 1987. He obtained his B.Sc (2012) and M.Sc (2014) in Telecommunication engineering at the University of Pisa. Since then he is pursuing a PhD at the University of Luxembourg on satellite communications with focus on link technologies for optical satellite feeder links. His research interests include Free Space Optics, Signal Processing, Satellite Systems and Random Access.

PLACE PHOTO HERE

**Riccardo De Gaudenzi** (M'89-SM"97) received his Doctor Engineer degree (cum Laude) in electronic engineering from the University of Pisa, Italy in 1985 and the PhD from the Technical University of Delft, The Netherlands in 1999. From 1986 to 1988 he was with the European Space Agency (ESA), Stations and Communications Engineering Department, Darmstadt (Germany) where he was involved in satellite Telemetry, Tracking and Control (TT&C) ground systems design and testing. In 1988, he joined ESAs Research and Technology Centre (ESTEC), Noordwijk, The Netherlands where since 2017 he is the Head of the ESAs Radio Frequency Systems and Payload Office. The Office, is responsible for supporting the definition and development of advanced satellite system, subsystems and related technologies for telecommunications, navigation and Earth observation applications. In 2016 he has been nominated ESA Competence Domain Lead for End-to-end RF/Optical systems and products for navigation, communication and remote sensing. In 1996 he spent one year with Qualcomm Inc., San Diego USA, in the Globalstar LEO project system group under an ESA fellowship. His current interest is mainly related with advanced telecommunication and navigation systems and techniques. More specifically, efficient digital modulation and multiple access techniques for fixed and mobile satellite services, synchronization topics, adaptive interference mitigation techniques and communication systems simulation techniques. He actively contributed to the development and the demonstration of the ETSI S-UMTS Family A, S-MIM, DVB-S2, DVB-S2X, DVB-RCS2 and DVB-SH standards. He has published more than 140 scientific papers and own more than 20 patents. He is co-recipient of the 2003 and 2008 Jack Neubauer Memorial Award Best Paper from the IEEE Vehicular Technology Society and of the IEEE Communication Society 2015 Satellite Communications Distinguished Service Award.

PLACE PHOTO HERE

**Čedomir Stefanović** (S'04-M'11-SM17) received his Dipl.-Ing., Mr.-Ing., and Ph.D. degrees in electrical engineering from the University of Novi Sad, Serbia. He is currently an associate professor at the Department of Electronic Systems, Aalborg University, Denmark. In 2014 he was awarded an individual postdoc grant by the Danish Council for Independent Research (Det Frie Forskningsrd). His research interests include communication theory, wireless and smart grid communications.